\documentclass[reprint,superscriptaddress,showpacs]{revtex4-1}
\usepackage{amsmath}
\usepackage{graphicx}
\usepackage{amssymb}

\begin{document}

\title{Noise correlation and success probability in coherent Ising machines}

\author{Yoshitaka Inui}
\email[]{yoshitaka.inui@ntt-research.com}

\affiliation{Physics and Informatics Laboratories, NTT Research Inc., 1950 University Ave., E. Palo Alto, CA 94303, USA}
\affiliation{National Institute of Informatics, Hitotsubashi 2-1-2, Chiyoda-ku, Tokyo 101-8430, Japan}

\author{Yoshihisa Yamamoto}

\affiliation{Physics and Informatics Laboratories, NTT Research Inc., 1950 University Ave., E. Palo Alto, CA 94303, USA}
\affiliation{E. L. Ginzton Laboratory, Stanford University, Stanford, CA 94305, USA}

\date{\today}

\begin{abstract}

We compared the noise correlation and the success probability of 
coherent Ising machines (CIMs) with optical delay-line, 
measurement feedback, and mean-field couplings. 
We theoretically studied three metrics for the noise correlations in these CIMs: quantum entanglement, quantum discord, 
and normalized correlation of canonical coordinates. 
The success probability was obtained through numerical simulations 
of truncated stochastic differential equations based on the Wigner distribution function. 
The results indicate that the success probability is more 
directly related to the normalized correlation function rather than entanglement or quantum discord. 

\end{abstract}


\maketitle

\section{Introduction}

Coherent Ising machines (CIMs) \cite{Wang13,Marandi14,Takata16,Inagaki16,McMahon16,Inagaki16b,Hamerly19} 
are dissipatively coupled networks of degenerate optical parametric oscillators (DOPOs), 
which can be used for solving combinatorial optimization problems and simulating various spin glass models. 
CIMs exploit the oscillation dynamics of a DOPO 
whose squeezed vacuum state below a threshold bifurcates into two displaced squeezed coherent states 
with positively and negatively valued mean amplitudes \cite{Wolinsky88,Kinsler91} above the threshold.   
Above the oscillation threshold, DOPOs in CIM tend to have a spin configuration 
determined by the mean amplitudes 
with the smallest loss for a given coupling matrix $\tilde{J}_{r,r'}$. 
When a CIM reaches the bifurcation threshold, where the parametric gain exceeds a linear loss, 
the correlation between the fluctuations in the DOPOs undergoes a significant surge. 
Such noise correlations at the threshold may determine the computational performance of the CIMs. 
However, it has remained unclear as to what types of correlation determine actual performance. 
Until now, the coupling matrices $\tilde{J}_{r,r'}$ have been introduced via an optical delay line \cite{Marandi14,Takata16,Inagaki16} 
or via homodyne measurement followed by coherent injection feedback \cite{McMahon16,Inagaki16b,Hamerly19}. 
It is unknown which scheme has a larger success probability or why the one has a better performance than the other. 

In this paper, we explain the computational performance of CIMs with noise correlations at the oscillation threshold. 
Each DOPO composing the CIM has a squeezed signal mode where the parametric interaction amplifies 
the canonical coordinate $\hat{X}=\frac{\hat{a}+\hat{a}^{\dagger}}{\sqrt{2}}$ 
and deamplifies the canonical momentum $\hat{P}=\frac{\hat{a}-\hat{a}^{\dagger}}{\sqrt{2}i}$. 
With an optical delay line (ODL) coupling, the squeezed/anti-squeezed vacuum noise of the $\hat{X}$/$\hat{P}$ components 
in one DOPO correlates with that of the other DOPO. 
If the coefficient of dissipative coupling is sufficiently large, 
a CIM satisfies the Duan-Giedke-Cirac-Zoller's sufficient criterion for entanglement \cite{Duan00,Takata15,Maruo16,Inui}. 
It also has non-zero quantum discord near its oscillation threshold \cite{Takata15,Inui}. 
Entanglement and quantum discord are correlation characteristics depending on both $\hat{X}$ and $\hat{P}$ components. 
Therefore, if both components contribute to the computational operation of the machine, 
entanglement or quantum discord may be useful metrics of performance. 
However, the computational performance of a CIM might depend only on the canonical coordinate $\hat{X}$. 
If so, the magnitude of entanglement and quantum discord would not be directly related to performance. 
Instead, a metric depending on only $\hat{X}$, for example, the noise correlation function of the $\hat{X}$ components, 
would be a good metric of the computational performance. 

We numerically studied the success probabilities of an optical delay-line coupled CIM (ODL-CIM) \cite{Marandi14,Takata16,Inagaki16}, 
a measurement feedback coupled CIM (MFB-CIM) \cite{McMahon16,Inagaki16b,Hamerly19}, 
and an ODL-CIM with a mean-field approximation for the coupling fields (MFA-CIM). 
We determined which metric most directly governs the success probabilities. 
We found that a normalized correlation function of 
the canonical coordinate $\hat{X}$ is the best metric that explains the behavior of the success probabilities. 
In some cases, this metric can predict the point at which 
the success probabilities of two CIMs with different coupling schemes cross. 
On the other hand, entanglement and quantum discord failed to explain the performance of the CIMs. 

In this paper, the success probability is calculated using the stochastic differential equations (SDEs), 
which after making appropriate truncations are equivalent to the density operator master equation and 
the Fokker-Planck equation for the Wigner distribution function. 
We present a theoretical model for an MFB-CIM in Sec. II. 
In Sec. III, we consider CIMs composed of two coupled DOPOs. 
For the ODL-CIM and MFA-CIM, we cite the corresponding results from the previous paper \cite{Inui}. 
We show that the normalized noise correlation function is the best metric of success probability for them. 
In Sec. IV, we present several modified models of two coupled DOPOs 
and show that their performance can be explained in terms of the normalized correlation function. 
In Sec. V, we consider a one-dimensional lattice of DOPOs and 
show that the normalized noise correlation function is also a useful metric for this extended system. 
Section VI summarizes the paper. 

\section{Theoretical Method}

\subsection{Stochastic differential equation for a solitary DOPO}

Here, we present the quantum model of a single DOPO. 
The quantum master equation for the density matrix $\hat{\rho}$ is 
\begin{equation}
\frac{\partial \hat{\rho}}{\partial t}=\mathcal{L}_{DOPO}\hat{\rho}=-\frac{i}{\hbar}[\hat{H},\hat{\rho}]+\sum_{j=1,2}([\hat{L}_j,\hat{\rho}\hat{L}_j^{\dagger}]+{\rm h.c.}), 
\end{equation}
where $\hat{H}=i\hbar \frac{p}{2}(\hat{a}^{\dagger 2}-\hat{a}^2)$ is the optical parametric interaction Hamiltonian, 
and $\hat{L}_1=\hat{a}$ and $\hat{L}_2=\sqrt{\frac{g^2}{2}}\hat{a}^2$ 
are the projectors for a linear loss and two-photon (nonlinear) loss. 
Here, the time $t$ is normalized in order that the linear loss rate is equal to one. 
The Wigner function is defined as the Weyl-ordered associative function of the density matrix \cite{Cahill69,Walls}:
\begin{equation}
W(\alpha)=\frac{1}{\pi^2}\int {\rm tr} \hat{\rho}e^{\eta(\hat{a}^{\dagger}-\alpha^*)-\eta^*(\hat{a}-\alpha)}d^2 \eta. 
\end{equation}
The Wigner expansion of the density matrix is \cite{Corney03}:
\begin{eqnarray}
\hat{\rho} &=& \int W(\alpha)\hat{\Lambda}_W(\alpha) d^2\alpha, \\
\hat{\Lambda}_W(\alpha) &=& 2\sum_{n=0}^{\infty}\frac{(-2)^n}{n!}(\hat{a}^{\dagger}-\alpha^*)^n(\hat{a}-\alpha)^n.
\end{eqnarray}
We can derive the Fokker-Planck equation for a single DOPO by using this expansion, the following relations, 
\begin{eqnarray}
\hat{a}\hat{\Lambda}_W &=& \Bigl(\alpha-\frac{1}{2}\frac{\partial}{\partial \alpha^*}\Bigr)\hat{\Lambda}_W, \\
\hat{a}^{\dagger}\hat{\Lambda}_W &=& \Bigl(\alpha^{*}+\frac{1}{2}\frac{\partial}{\partial \alpha}\Bigr)\hat{\Lambda}_W,
\end{eqnarray}
and partial integration. We assume a small saturation parameter ($g^2\ll 1$) and 
neglect the third-order derivatives in the Fokker-Planck equation. 
We derive the following SDE by using the Ito rule\cite{Wang13},
\begin{equation}
\label{wsde}
\frac{d\alpha}{dt}=-\alpha+p \alpha^*-g^2 |\alpha|^2\alpha+\sqrt{\frac{1}{2}+g^2 |\alpha|^2}\xi_C.
\end{equation}
Here $\xi_C$ is a complex random variable satisfying $\langle \xi_C^*(t)\xi_C(t')\rangle=2\delta(t-t')$. 

\subsection{Measurement-feedback circuit}

\begin{figure}[h]
\begin{center}
\includegraphics[width=6.0cm]{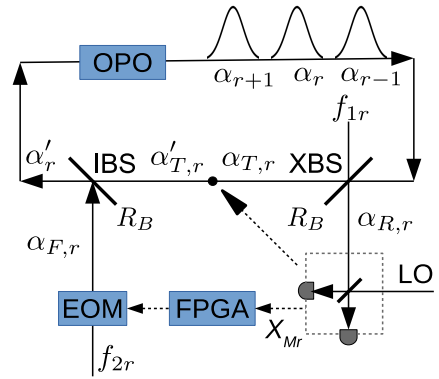}
\caption{Traveling DOPO pulse model of MFB-CIM.}
\label{model}
\end{center}
\end{figure}

A CIM consists of DOPOs denoted by $\hat{a}_r(r=1,\cdots,N)$ and a mutual coupling circuit. 
The total master equation is $\frac{\partial \hat{\rho}}{\partial t}=\sum_r\mathcal{L}_{DOPO}^{(r)}\hat{\rho}+\frac{\partial \hat{\rho}}{\partial t}|_C$, 
where $\mathcal{L}_{DOPO}^{(r)}$ operates only on the $r$-th DOPO, and $\frac{\partial \hat{\rho}}{\partial t}|_C$ describes the coupling between the DOPOs. 
Here, we describe the coupling projectors in an MFB-CIM. 
Fig. \ref{model} presents the coupling scheme for traveling DOPO pulses in the MFB-CIM. 
We assume that the roundtrip time of the ring cavity $\Delta t$ is sufficiently small 
compared with the unit time (linear loss induced decay time), i.e., $\Delta t\ll 1$. 
Therefore, the loss and gain per round trip is small, 
and a coarse-grained description of the machine in terms of SDEs is valid. 
Suppose that the extraction beam splitter (XBS) has reflectance $R_B=j\Delta t$, 
to the incident vacuum fluctuation noise $f_{1r}$ from an open port, 
then the reflected and transmitted amplitudes are, 
without loss of generality, $\alpha_{R,r}=\sqrt{R_B}\alpha_r-\sqrt{1-R_B}f_{1r}$, and 
$\alpha_{T,r}=\sqrt{1-R_B}\alpha_r+\sqrt{R_B}f_{1r}$. 
The random variables for the vacuum noise satisfy $\langle f_{ar}^*f_{br'}\rangle=\frac{1}{2}\delta_{ab}\delta_{rr'}$. 
When the real part of $\alpha_{R,r}$ is measured, 
the transmitted mode $\alpha_{T,r}$ receives the effect of the measurement and 
is converted to $\alpha_{T,r}'$ depending on the result of the measurement. 
This particular step describes a partial reduction of the internal DOPO pulse state 
by an indirect quantum measurement\cite{Braginsky}. 
The field programmable gate array (FPGA) calculates the amplitudes of the injection feedback signals from the measured values $X_{Mr}$. 
The injection feedback signals for the $r$-th pulse are expressed 
using a dimensionless matrix $\tilde{J}_{r,r'}$ as $\sum_{r'}\tilde{J}_{r,r'}X_{Mr'}$. 
The average amplitudes of the coherent injection fields are set 
to the calculated values by using an electrooptic modulator (EOM). 
The injection pulse also carries the vacuum noise ($f_{2r}$ in Fig. \ref{model}). 
The feedback injection amplitude is thus $\alpha_{F,r}=f_{2r}+\frac{1}{\sqrt{2}}\sum_{r'}\tilde{J}_{r,r'}X_{Mr'}$. 
The intracavity mode after the injection beam splitter (IBS) is 
$\alpha_r'=\sqrt{1-R_B}\alpha_{T,r}'+\sqrt{R_B}\alpha_{F,r}$. 

The coupling part $\frac{\partial \hat{\rho}}{\partial t}|_C$ includes the linear loss at the XBS represented by the Liouvillian
$\frac{j}{2} \sum_r([\hat{a}_r,\hat{\rho}\hat{a}_r^{\dagger}]+{\rm h.c.})$. 
The strength of the indirect homodyne measurement is related to the loss at the XBS. 
At the IBS, the transmitted mode experiences the same loss as at the XBS. 
The total loss of the cavity mode with two beam splitters can be written as $j \sum_r([\hat{a}_r,\hat{\rho}\hat{a}_r^{\dagger}]+{\rm h.c.})$.
$j$ is the total loss at the two beam splitters normalized by the linear background loss of a single DOPO.
This loss can be compensated for by using the coherent feedback injection, 
and $j$ works as the strength of the coupling between traveling pulses. 
We will choose the amplitudes of coherent feedback injection in a way that each DOPO pulse in the CIM 
has the same photon number as that of a solitary DOPO without two beam splitters. 

\subsection{Microscopic model of MFB-CIM}

Here, we will consider here a microscopic model of the MFB-CIM depicted in Fig.\ref{model}.  
In the MFB-CIM, only the $\hat{X}$ component is coupled, whereas the $\hat{P}$ component is left uncoupled. 
A measurement on $\alpha_{T,r}$ appears as a shift in the mean amplitude and a reduction in the fluctuation of 
${\rm Re}\alpha_r$ of the intracavity field (see also Appendix A).  
The feedback signal also operates on the real part of $\alpha_r$. 
Let us consider the equations of the real and imaginary parts, $X=\sqrt{2}{\rm Re}\alpha$ and $P=\sqrt{2}{\rm Im}\alpha$, 
instead of the complex amplitude (Eq. (\ref{wsde})): 
\begin{equation}
\frac{dX}{dt} = -(1-p)X-\frac{g^2}{2} (X^2+P^2)X +\sqrt{1+g^2(X^2+P^2)}\xi_{R1},
\end{equation}
\begin{equation}
\frac{dP}{dt} = -(1+p)P-\frac{g^2}{2} (X^2+P^2)P +\sqrt{1+g^2(X^2+P^2)}\xi_{R2}.
\end{equation}
Here, $\xi_{Ra}(a=1,2)$ are real-valued random variables satisfying $\langle \xi_{Ra}(t)\xi_{Rb}(t')\rangle=\delta_{ab}\delta(t-t')$. 
For measurement-feedback coupled DOPOs, the microscopic equations are as follows (see also Appendix A):
\begin{eqnarray}
\label{Xms}
\frac{dX_r}{dt} &=& -(1-p+j)X_r-\frac{g^2}{2} (X_r^2+P_r^2)X_r \nonumber \\
&+& \sqrt{1+j+g^2(X_r^2+P_r^2)-2j\langle :\Delta \hat{X}_r^2:\rangle^2}\xi_{R1,r} \\
&+& \sum_{r'} \tilde{J}_{rr'} \Bigl( j\langle X_{r'}\rangle+\sqrt{\frac{j}{2}}w_{R,r'}\Bigr) +\sqrt{2j}\langle :\Delta \hat{X}_r^2:\rangle w_{R,r},\nonumber \\
\label{Pms}
\frac{dP_r}{dt} &=& -(1+p+j)P_r-\frac{g^2}{2} (X_r^2+P_r^2)P_r \nonumber \\
&+& \sqrt{1+j+g^2(X_r^2+P_r^2)}\xi_{R2,r}. 
\end{eqnarray}
Here, $\langle \xi_{Ra,r}(t)\xi_{Rb,r'}(t')\rangle=\delta_{ab}\delta_{rr'}\delta(t-t')$.  
$\tilde{J}_{r,r'}$ is a dimensionless matrix representing the Ising coupling coefficients. 
$w_{R,r}$ is a real-valued random variable satisfying $\overline{w_{R,r}(t) w_{R,r'}(t')}=\delta_{r,r'}\delta(t-t')$, 
which accounts for the finite measurement error of the homodyne detector. 
The overline means the ensemble average. 
The $P$ component has no effect by measurement-feedback coupling except for the increased linear loss denoted by $j$. 
The $X$ component has a measurement-induced fluctuation reduction term, 
measurement-induced mean field shift term, and coherent injection term, 
i.e., the third, fifth, and fourth terms of the R.H.S. of Eq. (\ref{Xms}). 
Below, we refer to this model as the MFB-CIM (MI). 
In order to integrate Eq. (\ref{Xms}) numerically, we must evaluate the average amplitude $\langle X_r\rangle$ and 
normally ordered variance $\langle :\Delta \hat{X}_r^2: \rangle$ at each time step. 
To calculate these values, we simultaneously solve equations for many parallel DOPOs 
that are driven by identical measurement random variables $w_{R,r}$, 
but independent reservoir random variables $\xi_{Ra,r} (a=1,2)$. 
We will call a particle producing averaged values a 'Brownian particle'. 
The average over Brownian particles is taken to evaluate $\langle \hat{X}_r\rangle$ and $\langle : \Delta \hat{X}_r^2:\rangle$ at each time step. 

\subsection{Macroscopic model of MFB-CIM}

\begin{figure*}
\begin{center}
\includegraphics[width=14.0cm]{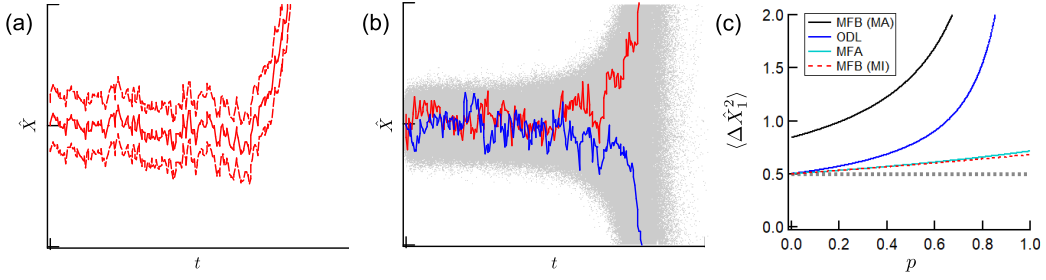}
\caption{Comparison of microscopic and macroscopic models of MFB-CIM. 
(a) Microscopic model. 
(b) Macroscopic model. 
(c) Steady-state fluctuations of two-site CIMs as a function of excitation $p$ with $j=7/3$. 
The vacuum fluctuation is shown by the gray dashed line. }
\label{ma}
\end{center}
\end{figure*}

Here, we present the macroscopic model derived in Ref. \cite{Wiseman93}, 
which is obtained by ensemble-averaging over many measurement records and is used in the study of the MFB-CIM \cite{Haribara15,Haribara17}. 
This model is not a microscopic model, but is rather considered to be a phenomenological model. 
The master equation for the coupling part consists of a measurement-induced state-reduction part 
\begin{eqnarray}
\label{qmema1}
\left.\frac{\partial \hat{\rho}}{\partial t}\right\vert_{C,sr} &=& 
\frac{j}{2} \sum_{r}([\hat{a}_r,\hat{\rho}\hat{a}_r^{\dagger}]+{\rm h.c.}) \\
&+& 2\sqrt{j}\sum_{r}\Bigl(\frac{\hat{a}_{r}\hat{\rho}+\hat{\rho}\hat{a}^{\dagger}_{r}}{2}-\frac{\langle \hat{a}_r+\hat{a}^{\dagger}_r\rangle}{2}\hat{\rho}\Bigr)w_{R,r}, \nonumber
\end{eqnarray}
and a coherent-injection part 
\begin{eqnarray}
\label{qmema2}
\left.\frac{\partial \hat{\rho}}{\partial t}\right\vert_{C,fb}&=&\frac{j}{2} \sum_{r}([\hat{a}_r,\hat{\rho}\hat{a}_r^{\dagger}]+{\rm h.c.}) \\
&+& j \sum_{r,r'}\tilde{J}_{r,r'}\Bigl(\frac{\langle \hat{a}_{r'}+\hat{a}^{\dagger}_{r'}\rangle}{2}+\frac{w_{R,r'}}{2\sqrt{j}}\Bigr) [\hat{a}_r^{\dagger}-\hat{a}_r,\hat{\rho}]. \nonumber 
\end{eqnarray}
Ensemble averaging over the noise variables $w_{R,r}$ in Eqs. (\ref{qmema1}) and (\ref{qmema2}) 
yields the following density matrix master equation: 
\begin{eqnarray}
\left.\frac{\partial \hat{\rho}}{\partial t}\right\vert_C &=& j\sum_{r}([\hat{a}_r,\hat{\rho}\hat{a}_r^{\dagger}]+{\rm h.c.})\nonumber \\
&+& \frac{j}{2}\sum_{r,r'}\tilde{J}_{r,r'}[\hat{a}_r^{\dagger}-\hat{a}_r,\hat{a}_{r'}\hat{\rho}+\hat{\rho}\hat{a}_{r'}^{\dagger}] \\
&+& \frac{j}{8}\sum_{r,r',r''}\tilde{J}_{r,r''}\tilde{J}_{r',r''}[\hat{a}_r^{\dagger}-\hat{a}_r,[\hat{a}_{r'}^{\dagger}-\hat{a}_{r'},\hat{\rho}]]. \nonumber
\end{eqnarray}
We can obtain the Fokker-Planck equation for the coupling part by using 
$\hat{a}\hat{\Lambda}_W+\hat{\Lambda}_W\hat{a}^{\dagger}=2{\rm Re} \alpha \hat{\Lambda}_W- {\rm Re}\frac{\partial}{\partial \alpha}\hat{\Lambda}_W$ and 
$[\hat{a}^{\dagger}-\hat{a},\hat{\Lambda}_W] =\Bigl(\frac{\partial}{\partial \alpha}+\frac{\partial}{\partial \alpha^*}\Bigr)\hat{\Lambda}_W$, 
\begin{eqnarray}
\left.\frac{\partial W}{\partial t}\right\vert_C &=& j\sum_{r}\Bigl(\frac{\partial}{\partial \alpha_r}(\alpha_r W)+\frac{\partial}{\partial \alpha_r^*}(\alpha_r^* W)+\frac{\partial^2 W}{\partial \alpha_r^* \partial \alpha_r}\Bigr) \nonumber \\
&-& \frac{j}{2}\sum_{r,r'}\tilde{J}_{r,r'}\Bigl(\frac{\partial}{\partial \alpha_r}+\frac{\partial}{\partial \alpha_r^{*}}\Bigr)(\alpha_{r'}+\alpha_{r'}^{*})W  \nonumber \\
&-& \frac{j}{4}\sum_{r,r'}\tilde{J}_{r,r'}\Bigl(\frac{\partial}{\partial \alpha_r}+\frac{\partial}{\partial \alpha_r^{*}}\Bigr)\Bigl(\frac{\partial}{\partial \alpha_{r'}}+\frac{\partial}{\partial \alpha_{r'}^{*}}\Bigr)W \\
&+& \frac{j}{8}\sum_{r,r',r''}\tilde{J}_{r,r''}\tilde{J}_{r',r''}\Bigl(\frac{\partial}{\partial \alpha_r}+\frac{\partial}{\partial \alpha_r^{*}}\Bigr)\Bigl(\frac{\partial}{\partial \alpha_{r'}}+\frac{\partial}{\partial \alpha_{r'}^{*}}\Bigr)W. \nonumber 
\end{eqnarray}
Next, we obtain the $c$-number SDE of the coupling part by following the Ito rule,  
\begin{eqnarray}
\label{sdema}
\left.\frac{d\alpha_r}{d t}\right\vert_C &=& -j\alpha_r-\sqrt{\frac{j}{4}}\xi_{C1,r} +\sqrt{\frac{j}{4}}\xi_{C2,r} \nonumber \\
&+& \sum_{r'}\tilde{J}_{r,r'}\Bigl(j{\rm Re}\alpha_{r'}+\sqrt{\frac{j}{4}}{\rm Re}\xi_{C1,r'}\Bigr). 
\end{eqnarray}
Here, $\langle \xi_{Ca,r}^*(t)\xi_{Cb,r'}(t')\rangle=2\delta_{ab}\delta_{rr'}\delta(t-t')$. 
Below, we refer to this model as the MFB-CIM (MA). 
In this theory, the observable of the 1st DOPO $\mathcal{O}_1$ and that of the 2nd DOPO $\mathcal{O}_2$ are generally correlated, 
$\langle \mathcal{O}_1 \mathcal{O}_2\rangle \ne \langle \mathcal{O}_1\rangle \langle \mathcal{O}_2\rangle$, 
by the ensemble averaging over $w_{R,r}$. 
The ideas behind the MFB-CIM (MA) model and MFB-CIM (MI) model are compared in Fig. \ref{ma}. 
Fig. \ref{ma} (a) illustrates the time ($t$) dependent dynamics of the mean field amplitude $\langle \hat{X}\rangle $ and 
variance $\langle \Delta \hat{X}^2\rangle$ in the MFB-CIM (MI). 
The red line shows the motion of the mean amplitude 
depending on the sequences of measurements. 
The dashed red lines show the range of quantum fluctuation around the mean amplitude. 
The macroscopic model describes many particles' motion in the configuration space within the gray area in Fig.\ref{ma}(b). 
These particles have larger fluctuations than the mean amplitude in Fig. \ref{ma} (a).  

\subsection{Gaussian model of MFB-CIM}

The microscopic model requires many Brownian particles with identical $w_{R,r}$ 
and independent $\xi_{Ra,r} (a=1,2)$ to be simulated even in a single run. 
As we have to consider the success probability for a sufficient number of simulation runs, 
the numerical cost of simulating the MFB-CIM (MI) model is higher than that of the MFB-CIM (MA) model. 
Here, we present a Gaussian approximation of the MFB-CIM (MI) model, following the idea in Ref. \cite{Shoji17}. 
We split the amplitude $\alpha_r$ into the mean amplitude $\langle \alpha_r\rangle$ and a small fluctuation $\Delta \alpha_r$, 
$\alpha_r =\langle \alpha_r \rangle+\Delta \alpha_r$. 
Then we solve equations of three real variables, $\mu_r =\langle \alpha_r \rangle=\langle \alpha_r^* \rangle$, 
$m_r =\langle \Delta \alpha_r^2\rangle=\langle \Delta \alpha_r^{*2}\rangle$, and 
$n_r =\langle | \Delta \alpha_r |^2 \rangle$, for each DOPO. 
The equations for the MFB-CIM (MI) are as follows:
\begin{eqnarray}
\frac{d\mu_r}{dt} &=& -(1-p+j)\mu_r-g^2 (\mu_r^2+2n_r+m_r)\mu_r \\
&+& \sum_{r'} \tilde{J}_{rr'} \Bigl( j \mu_{r'}+\sqrt{\frac{j}{4}}w_{R,r'}\Bigr) +\sqrt{j}\langle :\Delta \hat{X}_r^2:\rangle w_{R,r}, \nonumber \\
\frac{d n_r}{dt} &=& -2(1+j)n_r+2pm_r-2g^2\mu_r^2 (2n_r+m_r) \nonumber \\
&-& j \langle:\Delta \hat{X}_r^2:\rangle^2+1+j+2g^2(\mu_r^2+n_r), \\
\frac{d m_r}{dt} &=& -2(1+j)m_r+2pn_r-2g^2\mu_r^2 (2m_r+n_r) \nonumber \\ 
&-& j \langle:\Delta \hat{X}_r^2:\rangle^2.
\end{eqnarray}
Here, $\langle :\Delta \hat{X}_r^2:\rangle=n_r+m_r-\frac{1}{2}$. 
If we assume $g^2 n_r, g^2 m_r \ll 1$, the dynamics of $\langle \Delta \hat{P}_r^2\rangle=n_r-m_r$ are independent of 
those of $\mu_r$ and $\langle \Delta \hat{X}_r^2\rangle=n_r+m_r$. 
Accordingly, we get the following equations for $\mu_r$ and $V_r:=\langle \Delta \hat{X}_r^2\rangle$: 
\begin{eqnarray}
\label{ga1}
\frac{d\mu_r}{dt} &=& -(1-p+j)\mu_r-g^2 \mu_r^3 \\
&+& \sum_{r'} \tilde{J}_{rr'} \Bigl( j \mu_{r'}+\sqrt{\frac{j}{4}}w_{R,r'}\Bigr) +\sqrt{j}\Bigl(V_r-\frac{1}{2}\Bigr)w_{R,r}, \nonumber \\
\label{ga2}
\frac{d V_r}{dt} &=& -2(1-p+j)V_r-6g^2\mu_r^2 V_r \nonumber \\
&-& 2j \Bigl(V_r-\frac{1}{2}\Bigr)^2+1+j+2g^2\mu_r^2.
\end{eqnarray}
We will refer to the Gaussian-approximation model as the MFB-CIM (GA). 

\section{CIM with two DOPOs}

\subsection{ODL-CIM and MFA-CIM}

In this section, we consider the fluctuation characteristics and 
success probability of the simplest CIM consisting of just two DOPOs. 
Before considering the MFB-CIM, we summarize the results 
for the previously studied ODL-CIM and MFA-CIM \cite{Inui}. 
For the ODL-CIM, we consider a mediating cavity model \cite{Takata15} with the coupling Liouvillian 
\begin{equation}
\label{lcodl}
\left.\frac{\partial \hat{\rho}}{\partial t}\right \vert_C=j([\hat{a}_1-\hat{a}_2,\hat{\rho}(\hat{a}_1^{\dagger}-\hat{a}_2^{\dagger})]+{\rm h.c.}). 
\end{equation}
With this coupling, the steady-state fluctuations of ODL-CIM below the threshold can be derived as \cite{Inui} 
\begin{eqnarray}
\langle \Delta \hat{X}_1^2\rangle=\langle \Delta \hat{X}_2^2\rangle &=& \frac{1}{2}+\frac{(1-p+j)p}{2(1-p)(1-p+2j)}, \\
\langle \Delta \hat{X}_1 \Delta \hat{X}_2 \rangle &=& \frac{pj}{2(1-p)(1-p+2j)}, \\
\langle \Delta \hat{P}_1^2\rangle=\langle \Delta \hat{P}_2^2\rangle &=& \frac{1}{2}-\frac{(1+p+j)p}{2(1+p)(1+p+2j)}, \\
\langle \Delta \hat{P}_1 \Delta \hat{P}_2 \rangle &=& -\frac{pj}{2(1+p)(1+p+2j)}.
\end{eqnarray}
The fluctuations of the $\hat{X}$ components diverge at the threshold ($p\rightarrow 1$), 
whereas those of the $\hat{P}$ components remain finite.  

The mean-field approximation for the coupling term is obtained by making the replacement 
$[\hat{a}_1-\hat{a}_2,\hat{\rho}(\hat{a}_1^{\dagger}-\hat{a}_2^{\dagger})]\rightarrow 
[\hat{a}_1-\langle \hat{a}_2\rangle ,\hat{\rho}(\hat{a}_1^{\dagger}-\langle \hat{a}_2^{\dagger}\rangle)]
+[\langle \hat{a}_1\rangle-\hat{a}_2,\hat{\rho}(\langle \hat{a}_1^{\dagger}\rangle -\hat{a}_2^{\dagger})]$. 
With the mean-field coupling, the steady-state fluctuations of the CIM below the threshold are 
\begin{eqnarray}
\langle \Delta \hat{X}_1^2\rangle=\langle \Delta \hat{X}_2^2\rangle &=& \frac{1}{2}+\frac{p}{2(1-p+j)}, \\
\langle \Delta \hat{X}_1 \Delta \hat{X}_2 \rangle &=& 0, \\
\langle \Delta \hat{P}_1^2\rangle=\langle \Delta \hat{P}_2^2\rangle &=& \frac{1}{2}-\frac{p}{2(1+p+j)}, \\
\langle \Delta \hat{P}_1 \Delta \hat{P}_2 \rangle &=& 0.
\end{eqnarray}
All these values are finite at the threshold ($p\rightarrow 1$). 

\subsection{Steady-state fluctuations of MFB-CIM (MA)}

Here, we consider the steady-state fluctuation of an MFB-CIM consisting of two DOPOs 
described by a coupling matrix $\tilde{J}=\begin{bmatrix} 0 & 1\\ 1&0 \end{bmatrix}$. 
In the ensemble-averaged theory, i.e., in the case of the MFB-CIM (MA) model, the SDEs are as follows:
\begin{eqnarray}
\frac{d\alpha_1}{dt} &=& -(1+j)\alpha_1+j{\rm Re}\alpha_2 +p \alpha_1^*-g^2 |\alpha_1|^2\alpha_1\\
&+& \sqrt{\frac{j}{4}}{\rm Re}\xi_{C2}-\sqrt{\frac{j}{4}}\xi_{C1}+\sqrt{\frac{1}{2}+\frac{j}{4}+g^2 |\alpha_1|^2}\xi_{C3},\nonumber \\
\frac{d\alpha_2}{dt} &=& -(1+j)\alpha_2+j{\rm Re}\alpha_1 +p \alpha_2^*-g^2 |\alpha_2|^2\alpha_2\\
&+& \sqrt{\frac{j}{4}}{\rm Re}\xi_{C1}-\sqrt{\frac{j}{4}}\xi_{C2}+\sqrt{\frac{1}{2}+\frac{j}{4}+g^2 |\alpha_2|^2}\xi_{C4}.\nonumber
\end{eqnarray}
Here, $\langle \xi_{Ca}^*(t)\xi_{Cb}(t')\rangle=2\delta_{ab}\delta(t-t')$. 

We consider steady-state fluctuations below the threshold. 
First, we assume small $g$ so that we can ignore $g$ dependent terms; 
the equations for $X_1$ and $X_2$ become:
\begin{equation}
\frac{dX_1}{dt}=-(1-p+j)X_1+jX_2+\sqrt{\frac{j}{2}}(\xi_{R2}-\xi_{R1})+\sqrt{1+\frac{j}{2}}\xi_{R3},
\end{equation}
\begin{equation}
\frac{dX_2}{dt}=-(1-p+j)X_2+jX_1+\sqrt{\frac{j}{2}}(\xi_{R1}-\xi_{R2})+\sqrt{1+\frac{j}{2}}\xi_{R4}.
\end{equation}
Here, $\langle \xi_{Ra}(t)\xi_{Rb}(t')\rangle=\delta_{ab}\delta(t-t')$. 
From these equations, the following equations for $\langle X_1^2\rangle$ and $\langle X_1X_2\rangle$ 
are obtained by assuming $\langle X_1^2\rangle=\langle X_2^2\rangle$: 
\begin{equation}
\frac{d\langle X_1^2\rangle}{dt}=-2(1-p+j)\langle X_1^2\rangle +2j\langle X_1 X_2\rangle +1+\frac{3}{2}j,
\end{equation}
\begin{equation}
\frac{d\langle X_1 X_2\rangle}{dt}=-2(1-p+j)\langle X_1 X_2\rangle +2j\langle X_1^2\rangle -j.
\end{equation}
The steady-state fluctuations are 
\begin{eqnarray}
\label{scma}
\langle \Delta \hat{X}_1^2\rangle = \langle \Delta \hat{X}_2^2\rangle &=& \frac{1}{2}+\frac{(1-p+j)(p+\frac{j}{2})}{2(1-p)(1-p+2j)}, \\
\label{ccma}
\langle \Delta \hat{X}_1 \Delta \hat{X}_2 \rangle &=& \frac{(p+\frac{j}{2})j}{2(1-p)(1-p+2j)}.
\end{eqnarray}
For the $\hat{X}$ components, these ensemble-averaged fluctuations of the MFB-CIM are larger than those of the ODL-CIM, 
whereas the $\hat{P}$ components of the MFB-CIM have the same characteristics as those of the MFA-CIM. 

\subsection{Steady-state fluctuations of MFB-CIM (MI)}

Here, we derive the steady-state fluctuations from the MFB-CIM (MI) model and 
show that the fluctuation characteristics produced by this model are identical to those of the MFB-CIM (MA) model. 
In the MFB-CIM (MI) model, Eq. (\ref{Xms}) satisfies 
$\langle \Delta \hat{X}_1\Delta \hat{X}_2\rangle=\langle \Delta \hat{X}_1\rangle \langle \Delta \hat{X}_2\rangle$.  
Here, we consider steady-state fluctuations of $\hat{X}$ below the threshold given by 
Eq.(\ref{Xms}) with $g\ll 1$,
\begin{eqnarray}
\label{Xmsbt}
\frac{dX_1}{dt} &=& -(1-p+j)X_1+ j\langle X_2\rangle+\sqrt{\frac{j}{2}}w_{R2}\\
&+& \sqrt{2j}\langle :\Delta \hat{X}_1^2:\rangle w_{R1} +\sqrt{1+j-2j\langle :\Delta \hat{X}_1^2:\rangle^2}\xi_{R1},\nonumber \\ 
\frac{dX_2}{dt} &=& -(1-p+j)X_2+ j\langle X_1\rangle+\sqrt{\frac{j}{2}}w_{R1}\\
&+& \sqrt{2j}\langle :\Delta \hat{X}_2^2:\rangle w_{R2} +\sqrt{1+j-2j\langle :\Delta \hat{X}_2^2:\rangle^2}\xi_{R2}.\nonumber 
\end{eqnarray}
Here, $\langle \xi_{Rr}(t)\xi_{Rr'}(t')\rangle=\delta_{rr'}\delta(t-t')$ are real random variables 
representing the quantum fluctuations from reservoirs. 
$\overline{w_{Rr}(t)w_{Rr'}(t')}=\delta_{rr'}\delta(t-t')$ are real random numbers 
representing the random deviation of a measurement result from the mean amplitude. 
The overline represents the ensemble average. 

First, let us consider the fluctuation $\xi_{Ra} (a=1,2)$ related to the reservoir noise, 
before taking the ensemble average over the measurement randomness $w_{R,r} (r=1,2)$. 
From Eq. (\ref{Xmsbt}), we obtain 
\begin{eqnarray}
\label{x1ms}
\frac{d\langle X_1\rangle}{dt} &=& -(1-p+j)\langle X_1\rangle+ j\langle X_2 \rangle \nonumber \\
&+& \sqrt{\frac{j}{2}} w_{R2}+\sqrt{2j}\langle:\Delta \hat{X}_1^2:\rangle w_{R1}, \\
\frac{d\langle X_1\rangle^2}{dt} &=& -2(1-p+j)\langle X_1\rangle^2+ 2j\langle X_1\rangle \langle X_2 \rangle \\
&+& \sqrt{2j}\langle X_1\rangle w_{R2}+2\sqrt{2j}\langle:\Delta \hat{X}_1^2:\rangle \langle X_1\rangle w_{R1}, \nonumber \\ 
\frac{d\langle X_1^2\rangle}{dt} &=& -2(1-p+j)\langle X_1^2\rangle+ 2j\langle X_1\rangle \langle X_2 \rangle\nonumber \\
&+& \sqrt{2j}\langle X_1\rangle w_{R2}+2\sqrt{2j}\langle:\Delta \hat{X}_1^2:\rangle \langle X_1\rangle w_{R1}\nonumber \\
&+& 1+j-2j\langle:\Delta \hat{X}_1^2:\rangle^2.
\end{eqnarray}
Here, the normally ordered fluctuation $\langle:\Delta \hat{X}_1^2:\rangle=\langle X_1^2\rangle-\langle X_1\rangle^2-\frac{1}{2}$ satisfies 
\begin{equation}
\frac{d\langle:\Delta \hat{X}_1^2:\rangle}{dt}=-2(1-p+j)\langle:\Delta \hat{X}_1^2:\rangle+p-2j\langle:\Delta\hat{X}_1^2:\rangle^2.
\end{equation}
Therefore, the steady-state fluctuation below the threshold is 
\begin{equation}
\label{dx2msmfb}
\langle \Delta \hat{X}_1^2\rangle=\frac{1}{2}+\frac{-(1-p+j)+\sqrt{(1-p+j)^2+2pj}}{2j}.
\end{equation}
This fluctuation is slightly smaller than that of the MFA-CIM because of the state reduction due to a homodyne-measurement. 
This value is not affected by the random sequences of the measurement results, 
at least below the oscillation threshold. 

Next, we will consider ensemble averaging over the measurement results. 
The expectation value for $\hat{X}_1$ 
after averaging over the quantum noise and ensemble-averaging over the measurement results is 
denoted as $\overline{\langle \hat{X}_1\rangle}$. 
$\overline{\langle \hat{X}_1\rangle}$ is zero below the threshold. 
We define the fluctuation around this value as $\overline{\Delta}\hat{X}_r:=\hat{X}_r-\overline{\langle \hat{X}_r\rangle}$. 
Consider the ensemble-averaged cross-correlation 
$\overline{\langle \overline{\Delta} \hat{X}_1\overline{\Delta}\hat{X}_2\rangle}$. 
This is calculated as $\overline{\langle \hat{X}_1\rangle \langle \hat{X}_2\rangle}$, assuming that two DOPOs are separable 
and that $\overline{\langle \hat{X}_r\rangle}=0$ below the threshold. 
From Eq. (\ref{x1ms}), we obtain 
\begin{eqnarray}
\frac{d\overline{\langle X_1\rangle^2}}{dt} &=& -2(1-p+j)\overline{\langle X_1\rangle^2}+ 2j\overline{\langle X_1 \rangle\langle X_2 \rangle} \nonumber \\
&+& \frac{j}{2}+2j\langle :\Delta \hat{X}_1^2:\rangle^2, \\
\frac{d\overline{\langle X_1 \rangle\langle X_2 \rangle}}{dt} &=& -2(1-p+j)\overline{\langle X_1 \rangle\langle X_2 \rangle}+ 2j\overline{\langle X_1\rangle^2} \nonumber \\
&+& 2j\langle :\Delta \hat{X}_1^2:\rangle.
\end{eqnarray}
From these equations and Eq. (\ref{dx2msmfb}), we obtain the ensemble-averaged correlation function, 
\begin{equation}
\overline{\langle \overline{\Delta} \hat{X}_1 \rangle \langle \overline{\Delta} \hat{X}_2 \rangle}=\frac{(p+\frac{j}{2})j}{2(1-p)(1-p+2j)}.
\end{equation}
This is identical to $\langle \Delta\hat{X}_1\Delta \hat{X}_2\rangle$ in Eq. (\ref{ccma}) in the macroscopic theory. 
On the other hand, 
\begin{equation}
\overline{\langle \overline{\Delta} \hat{X}_1\rangle^2}=\frac{(1-p+j)(p+\frac{j}{2})}{2(1-p)(1-p+2j)}-\langle :\Delta \hat{X}_1^2:\rangle
\end{equation}
is smaller than $\langle \Delta \hat{X}_1^2\rangle$ in the macroscopic theory. 
However, $\overline{\langle \hat{X}_1\rangle^2}=\overline{\langle \overline{\Delta} \hat{X}_1\rangle^2}$ differs from 
$\overline{\langle \hat{X}_1^2\rangle}=\overline{\langle \overline{\Delta} \hat{X}_1^2\rangle}$. 
Since $\langle \hat{X}_1^2\rangle=\langle \hat{X}_1\rangle^2+\frac{1}{2}+\langle :\Delta \hat{X}_1^2:\rangle$, 
\begin{equation}
\overline{\langle \overline{\Delta} \hat{X}_1^2\rangle}=\frac{1}{2}+\frac{(1-p+j)(p+\frac{j}{2})}{2(1-p)(1-p+2j)}
\end{equation}
is the same as $\langle \Delta \hat{X}_1^2\rangle$ in the macroscopic theory. 

Single-site fluctuations are summarized in Fig. \ref{ma} (c). 
The fluctuation of the MFB-CIM (MI) model is the smallest. 
It is slightly smaller than the fluctuation of the MFA-CIM model, 
because of the state reduction caused by the homodyne measurement.  
These two fluctuations don't have singular increases at the threshold. 
On the other hand, the fluctuations of the MFB-CIM (MA) and ODL-CIM models 
have singular increases at the threshold.  

\subsection{Metrics of noise correlation}

\begin{figure*}
\begin{center}
\includegraphics[width=14.0cm]{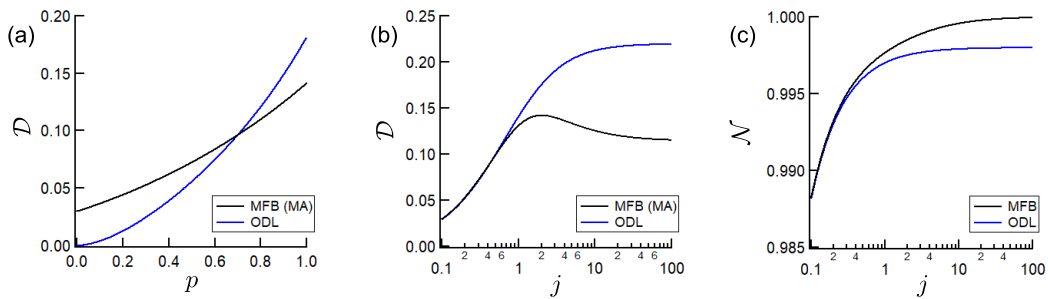}
\caption{Quantum discord $\mathcal{D}$ and normalized correlation function $\mathcal{N}$. (a) Quantum discord with $j=7/3$ as a function of excitation $p$. 
(b) Quantum discord with $p=0.999$ as a function of coupling coefficient $j$. 
(c) Normalized correlation function with $p=0.999$ as a function of coupling coefficient $j$. }
\label{qd}
\end{center}
\end{figure*}

Entanglement and quantum discord have been calculated for ODL-CIM \cite{Takata15,Maruo16}. 
For two coupled DOPOs, Duan's necessary and sufficient condition for entanglement \cite{Duan00} is 
satisfied for $j > \frac{1}{2}$ at the threshold \cite{Inui}. 
If $j$ is smaller than $1/2$, the entanglement criterion ceases to be satisfied before reaching the threshold. 
Quantum discord is calculated using the covariance matrix for the vector $\overrightarrow{\hat{R}}=\sqrt{2}[\hat{X}_1,\hat{P}_1,\hat{X}_2,\hat{P}_2]$. 
This covariance matrix has four non-zero independent values 
$a_1=2\langle \Delta \hat{X}_1^2\rangle=2\langle \Delta \hat{X}_2^2\rangle$, 
$a_2=2\langle \Delta \hat{P}_1^2\rangle=2\langle \Delta \hat{P}_2^2\rangle$, 
$c_1=2\langle \Delta \hat{X}_1 \Delta \hat{X}_2 \rangle$, 
and $c_2=2\langle \Delta \hat{P}_1 \Delta \hat{P}_2 \rangle$. 
If 
\begin{equation}
\label{qdcond}
(a_2c_1^2-a_1c_2^2(a_1^2-c_1^2))(a_2c_1^2(a_2^2-c_2^2)-a_1c_2^2)\ge 0,
\end{equation}
quantum discord is calculated as \cite{Giorda10,Adesso10} 
\begin{equation}
\mathcal{D}=f(\sqrt{a_1a_2})+f\Bigl(\sqrt{\frac{a_2}{a_1}(a_1^2-c_1^2)}\Bigr)-f(\nu_-)-f(\nu_+),
\end{equation}
where $f(x)=\frac{x+1}{2}\log\frac{x+1}{2}-\frac{x-1}{2}\log\frac{x-1}{2}$, 
and $\nu_{\pm}^2=(a_1\pm c_1)(a_2\pm c_2)$. 
Quantum discord contains the correlation of $\hat{P}$ as well as that of $\hat{X}$. 
Although it is zero for a density operator governing 
a single history of measurement results in the MFB-CIM (MI) model, 
nonzero quantum discord can be calculated in the density operator after the ensemble average is taken over many measurement records. 
If the canonical momentum contributes to the computational performance of CIMs, 
quantum discord could be a useful metric. 
Since $c_2=0$, the condition (\ref{qdcond}) is easily proven to be satisfied for the MFB-CIM (MA) model. 
Fig. \ref{qd} (a) presents the quantum discord for the ODL-CIM \cite{Inui} and MFB-CIM (MA) models, 
with $j=7/3$ as a function of the excitation $p$. 
Far below the threshold, quantum discord is larger for the MFB-CIM (MA) model, 
but at the threshold where the bifurcation happens, it is larger for the ODL-CIM model. 
Fig. \ref{qd} (b) plots the quantum discord near the threshold $p=0.999$ as a function of the coupling $j$. 
For smaller $j$, quantum discord is slightly larger for the MFB-CIM (MA) model, 
but for $j$ larger than $0.455$, it is larger for the ODL-CIM model. 
When the ODL-CIM model satisfies the entanglement criterion at the threshold, 
it always has a larger quantum discord than that of MFB-CIM (MA). 
The MFA-CIM model always has zero quantum discord. 
For the MFB-CIM (MA) model, the quantum discord at the threshold ($p\rightarrow 1$) with $j\rightarrow \infty$ is 
$\mathcal{D}\rightarrow f(\sqrt{\frac{5}{2}})-f(\sqrt{\frac{5}{4}})+\frac{1}{2}\log \frac{1}{2}\sim 0.114$.
This is smaller than the quantum discord of the ODL-CIM model $\mathcal{D}\rightarrow f(\sqrt{\frac{3}{2}})+\frac{1}{2}\log \frac{3}{4}\sim 0.220$, at the same limit. 

Next, let us consider a metric depending only on the $\hat{X}$ component: 
the normalized correlation function. Just below the threshold, we have 
\begin{equation}
\label{nc}
\mathcal{N}=\lim_{p\rightarrow 1-\delta}\frac{\overline{\langle \overline{\Delta} \hat{X}_1 \overline{\Delta} \hat{X}_2\rangle}}{\bigl(\overline{\langle \overline{\Delta} \hat{X}_1^2 \rangle}\cdot  \overline{\langle \overline{\Delta} \hat{X}_2^2\rangle}\bigr)^{\frac{1}{2}}}.
\end{equation}
Equation (\ref{nc}) becomes 
\begin{equation}
\mathcal{N}=1-\Big(\frac{1}{j}+2\Bigr)\delta+O(\delta^2) 
\end{equation}
for the ODL-CIM model, and 
\begin{equation}
\mathcal{N}=1-\Bigl(\frac{1}{j}+\frac{4}{2+j}\Bigr)\delta+O(\delta^2) 
\end{equation}
for the MFB-CIM model. 
The MFA-CIM model has $\mathcal{N}=0$. 
For the same $\delta$, the MFB-CIM model has a larger normalized correlation function $\mathcal{N}$ than that of the ODL-CIM model (Fig. \ref{qd} (c)). 
Table \ref{tab1} summarizes the steady-state characteristics of the three metrics with $j>1/2$ and $p\rightarrow 1$. 

\begin{table*}
\begin{center}
\caption{Three steady-state metrics with $j>1/2$ and $p\rightarrow 1$. }
\label{tab1}
\begin{tabular}{|c|c|c|c|}
\hline
 & Entanglement & Quantum Discord & Normalized correlation function of $\hat{X}$\\
\hline
MFB-CIM & No & Zero (MI) / Small (MA) & Large \\
ODL-CIM & Yes & Large & Small \\
MFA-CIM & No & Zero & Zero \\
\hline
\end{tabular}
\end{center}
\end{table*}

\subsection{Success probability}

Here, we describe the results of numerical simulations comparing the success probabilities $P_{sc}$ of the various CIM models. 
We simulated the time development from an initial vacuum state $\hat{\rho}=|0\rangle \langle 0|$ 
and judged whether the run was a success or failure at time $t=10$. 
We considered the parametric excitation $p$ depending on time 
\begin{equation}
p(t)=0.8+\frac{0.4}{e^{-(t-5)}+1}. 
\end{equation}
We set $g^2=10^{-4}$ and $\Delta t=2\times 10^{-3}$. 
For the ODL-CIM and MFB-CIM (MA) models, we simulated the time development of a single Brownian particle per DOPO. 
In the final time step, we compared the sign of ${\rm Re}\alpha_1$ with that of ${\rm Re}\alpha_2$. 
The simulation run was a success if the two signs were the same. 
The success probability $P_{sc}$ is defined as 
the number of successful runs divided by the total number of runs. 
We simulated $10^6$ runs on the ODL-CIM and MFB-CIM (MA) models. 
In the case of the MFA-CIM model, there were $10^2$ or $10^3$ Brownian particles per DOPO. 
For the MFB-CIM (MI) model, there were $10^4$ Brownian particles per DOPO. 
In these simulations, we randomly selected a single particle per DOPO and judged the success or failure. 
We performed $10^4$ runs for the MFA-CIM and MFB-CIM (MI) models. 
In Gaussian-approximation (GA) MFB-CIM, we calculated the time development of three values ($\mu_r$, $n_r$ and $m_r$) per DOPO. 
We performed $10^6$ runs for MFB-CIM (GA). 
At the final time step of the MFB-CIM (GA), we computed the Wigner amplitude for each DOPO 
as $\langle \hat{X}_r\rangle+\sqrt{\langle \Delta \hat{X}_r^2\rangle}N_r$. 
Here, $N_r$ is a normal random variable. 

\begin{figure*}
\begin{center}
\includegraphics[width=14.0cm]{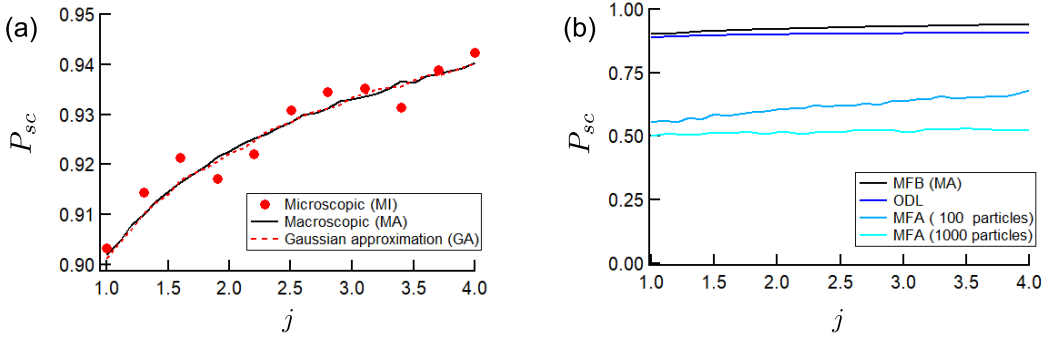}
\caption{Numerical success probability $P_{sc}$ of CIM with two DOPOs as a function of coupling coefficient $j$. 
(a) $P_{sc}$ of three MFB-CIM methods. (b) $P_{sc}$ of MFB-CIM, ODL-CIM, and MFA-CIM.}
\label{sp2s}
\end{center}
\end{figure*}

Fig. \ref{sp2s} (a) compares the results of the three MFB-CIM models (MI, MA, and GA). 
These models had almost identical success probabilities, 
although MFB-CIM (MI) was much more difficult to simulate. 
The macroscopic model and Gaussian-approximation model worked well as alternative methods. 
The dependence on $g^2$ is discussed in Appendix B. 
Fig. \ref{sp2s} (b) compares the results of the MFB-CIM (MA), ODL-CIM and MFA-CIM models (for $10^2$ or $10^3$ Brownian particles). 
The MFB-CIM model had the highest success probability, although the ODL-CIM model operated in the region where 
the entanglement criterion is satisfied and has a larger quantum discord than that of the MFB-CIM (MA) model at the threshold. 
The success probability of the MFA-CIM model in the case of $10^3$ Brownian particles 
was close to that of random-guess ($P_{sc}=\frac{1}{2}$). 
These results indicate that the success probability can be explained 
in terms of the magnitude of the normalized correlation function $\mathcal{N}$, 
rather than entanglement or quantum discord. 

\section{Discussion}

\subsection{Impact of state reduction caused by measurement}

\begin{figure*}
\begin{center}
\includegraphics[width=14.0cm]{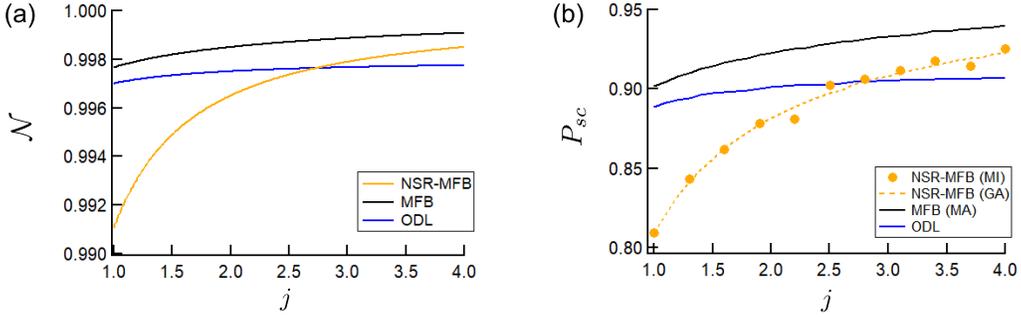}
\caption{Characteristics of two-site MFB-CIM with no state reduction (NSR-MFB-CIM) as a function of coupling coefficient $j$. 
(a) Normalized correlation function of NSR-MFB-CIM with $p=0.999$. 
(b) Numerical success probability $P_{sc}$ of NSR-MFB-CIM, MFB-CIM, and ODL-CIM.}
\label{disc1}
\end{center}
\end{figure*}

The MFB-CIM (MI) model incorporates the state reduction caused by optical homodyne measurement. 
Here, we will examine a model without a state reduction in order to explain the role of measurements in the MFB-CIM model. 
In this model, the equation for $X$ is modified as follows: 
\begin{eqnarray}
\frac{dX_r}{dt}&=&-(1-p+j)X_r +\sqrt{1+j+g^2(X_r^2+P_r^2)}\xi_{Rr}\\
&-& \frac{g^2}{2} (X_r^2+P_r^2)X_r+\sum_{r'} \tilde{J}_{rr'} \Bigl( j\langle X_{r'}\rangle+\sqrt{\frac{j}{2}}w_{Rr'}\Bigr). \nonumber 
\end{eqnarray}
Here, the measurement-induced mean amplitude shift and noise reduction are both absent. 
However, the randomness of the measured value remains in the equation. 
Let us use this equation to model two coupled DOPOs. 
We'll call it NSR-MFB-CIM. 
In this model, before taking the ensemble average, the $\hat{X}$ fluctuation is the same as that of the MFA-CIM model: 
$\langle \Delta \hat{X}_1^2\rangle=\frac{1}{2}+\frac{p}{2(1-p+j)}$. 
The ensemble-averaged fluctuations satisfy: 
\begin{eqnarray}
&&\overline{\langle \overline{\Delta} \hat{X}_1 \rangle \langle \overline{\Delta} \hat{X}_2 \rangle}=\frac{j^2}{4(1-p)(1-p+2j)}, \\
&&\overline{\langle \overline{\Delta} \hat{X}_1^2\rangle}=\frac{1}{2}+\frac{p}{2(1-p+j)}+\frac{(1-p+j)j}{4(1-p)(1-p+2j)}. 
\end{eqnarray}
The normalized correlation function is  
\begin{equation}
\mathcal{N}=1-\Bigl(\frac{1}{j}+\frac{4(1+j)}{j^2}\Bigr)\delta+O(\delta^2). 
\end{equation}
The normalized correlation function of the NSR-MFB-CIM model with $p=0.999$ is plotted 
together with those of the ODL-CIM and MFB-CIM models in Fig. \ref{disc1} (a). 
The NSR-MFB-CIM model always has a smaller $\mathcal{N}$ than that of the MFB-CIM model with state reduction. 
Moreover, it has an even smaller $\mathcal{N}$ than that of the ODL-CIM model for small $j$. 
In the case of $j=2$, the unnormalized correlation function 
$\overline{\langle \overline{\Delta} \hat{X}_1 \rangle \langle \overline{\Delta} \hat{X}_2 \rangle}$ 
of the NSR-MFB-CIM model is identical to that of the ODL-CIM model. 
However $j>1+\sqrt{3}$ is required for $\mathcal{N}$ of the NSR-MFB-CIM model to exceed that of the ODL-CIM model 
because of the larger $\overline{\langle \overline{\Delta} \hat{X}_1^2\rangle}$. 
The success probability of the NSR-MFB-CIM model is plotted in Fig. \ref{disc1} (b). 
It has the same value as the ODL-CIM model around $j\sim 2.8$. 
This supports our previous remark that 
the success probability is more directly related to the normalized correlation function $\mathcal{N}$. 
From these results, we can conclude the correlation between $\alpha_{T,r}$ and $\alpha_{R,r}$ 
and the partial state reduction by the homodyne measurement 
plays an essential role in the MFB-CIM model outperforming the ODL-CIM model particularly for small $j$. 

\subsection{Impact of thermal noise on ODL-CIM}

\begin{figure*}
\begin{center}
\includegraphics[width=18.0cm]{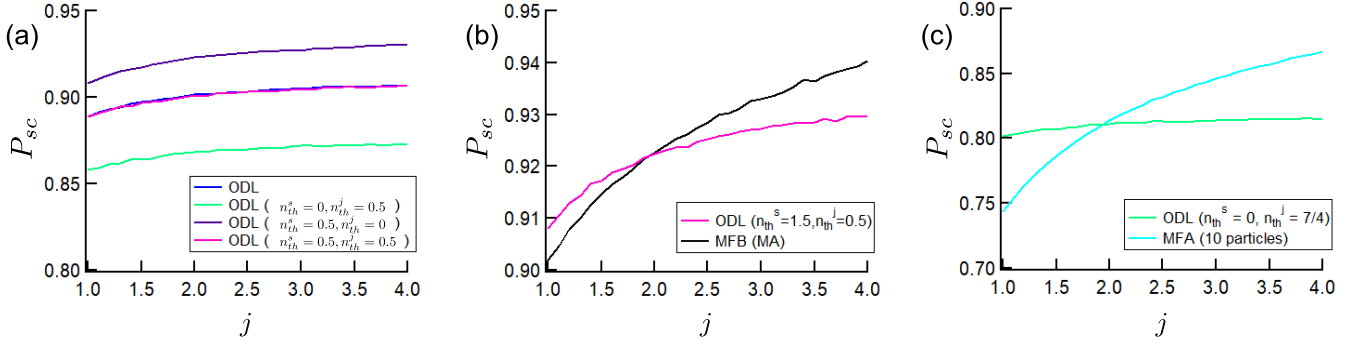}
\caption{Numerical success probability $P_{sc}$ of two-site ODL-CIM with thermal noise as a function of coupling coefficient $j$. 
(a) $P_{sc}$ of ODL-CIM with several thermal photon numbers. 
(b) $P_{sc}$ of ODL-CIM with thermal noise and MFB-CIM. 
(c) $P_{sc}$ of ODL-CIM with thermal noise and MFA-CIM.}
\label{disc2}
\end{center}
\end{figure*}

Here, we present examples of the close relation 
between the normalized correlation function $\mathcal{N}$ and success probability $P_{sc}$. 
First, we consider the impact of thermal noise in the ODL-CIM model. 
This modification introduces the following Liouvillian: 
\begin{eqnarray}
\label{lvth}
\left.\frac{\partial \hat{\rho}}{\partial t}\right \vert_{C,th} &=& 2n_{th}^s[\hat{a}_1,[\hat{\rho},\hat{a}_1^{\dagger}]]
+2n_{th}^s[\hat{a}_2,[\hat{\rho},\hat{a}_2^{\dagger}]]\nonumber \\
&+& 2jn_{th}^j[\hat{a}_1-\hat{a}_2,[\hat{\rho},\hat{a}_1^{\dagger}-\hat{a}_2^{\dagger}]]. 
\end{eqnarray}
Here, $n_{th}^s$ is thermal photon number of a reservoir mode 
which is responsible for the single mode loss, 
and $n_{th}^j$ is thermal photon number related to the dissipative coupling [Eq. (\ref{lcodl})]. 
The fluctuations of $\hat{X}$ are represented as follows:
\begin{equation}
\langle \Delta \hat{X}_1^2\rangle=\frac{1}{2}+\frac{(1-p+j)(p+2n_{th}^s)+2j(1-p)n_{th}^j}{2(1-p)(1-p+2j)},
\end{equation}
\begin{equation}
\langle \Delta \hat{X}_1 \Delta \hat{X}_2 \rangle=\frac{j(p+2n_{th}^s-2(1-p)n_{th}^j)}{2(1-p)(1-p+2j)}.
\end{equation}
The normalized correlation function is 
\begin{equation}
\mathcal{N}=1-\Bigl(\frac{1}{j}+\frac{2(1+2n_{th}^j)}{1+2n_{th}^s}\Bigr)\delta+O(\delta^2). 
\end{equation}
From this equation, when $n_{th}^j>0$ and $n_{th}^s=0$, 
the normalized correlation function becomes smaller and this would 
lead to a lower success probability, as was pointed out in Ref. \cite{Maruo16}. 
On the other hand, when $n_{th}^s>0$ and $n_{th}^j=0$, the normalized correlation function becomes larger, 
which would increase the success probability. 
When $n_{th}^s=n_{th}^j>0$, the normalized correlation function is the same as that with no thermal noise. 
Fig. \ref{disc2} (a) presents the numerical success probabilities of ODL-CIM with thermal noise.  
In the case of $n_{th}^s=0.5$ ($n_{th}^j=0.5$), the success probability increases (decreases), 
relative to the value of the model without thermal noise. 
In the case of $n_{th}^s=n_{th}^j=0.5$, the success probability is almost the same as 
that of the model without thermal noise. 
Next, we simulated the case of $n_{th}^s=1.5$ and $n_{th}^j=0.5$ and compared it with the MFB-CIM model. 
As shown in Fig. \ref{disc2} (b), the success probabilities coincide around $j\sim 1.93$, 
which is close to that predicted by the normalized correlation function $\mathcal{N}$ (at $j=2$). 
For smaller $j$, the ODL-CIM model with thermal noise has a higher success probability than that of the MFB-CIM model. 

Next, we compare the thermally injected ODL-CIM model ($n_{th}^j>0$) with 
the MFA-CIM model simulated by finite Brownian particles. 
The fluctuations of the MFA-CIM model depending on the 
number of Brownian particles ($N_p$) is expressed as
\begin{eqnarray}
\langle \Delta \hat{X}_1^2\rangle &=& \frac{1}{2}+\frac{p}{2(1-p+j)} \nonumber \\ 
&+&\frac{j^2(1+j)}{2N_p(1-p)(1-p+2j)(1-p+j)} \\
\langle \Delta \hat{X}_1\Delta \hat{X}_2\rangle &=& \frac{j(1+j)}{2N_p(1-p)(1-p+2j)}.
\end{eqnarray}
These values are larger than those of the positive-$P$ MFA-CIM model with the same $N_p$ \cite{Inui}, 
because Wigner theory has larger fluctuations than positive-$P$ theory. 
Consequently, the normalized correlation function is 
\begin{equation}
\mathcal{N}=1-\Bigl( \frac{1}{j}+\frac{2(N_p-1)}{j}\Bigr)\delta+O(\delta^2).
\end{equation}
This normalized correlation function with $N_p=10$ crosses the function of the ODL-CIM model with thermal noise $n_{th}^s=0,n_{th}^j=7/4$ 
at $j=2$. As shown in Fig. \ref{disc2} (c), the success probabilities of these two CIMs coincide around $j\sim 1.94$. 
This similarity supports the conclusion that the success probability is strongly related to the normalized correlation function. 

\subsection{Impact of squeezed reservoir on ODL-CIM}

\begin{figure}
\begin{center}
\includegraphics[width=7.0cm]{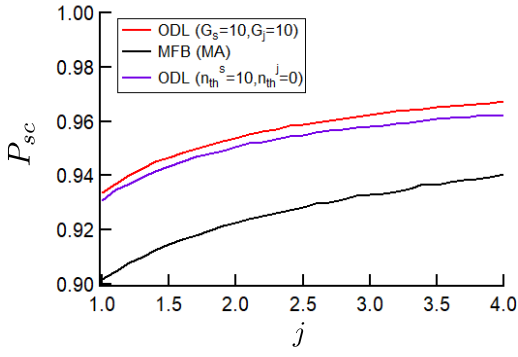}
\caption{Numerical success probability $P_{sc}$ of two-site ODL-CIM with squeezed noise as a function of coupling coefficient $j$. }
\label{disc3}
\end{center}
\end{figure}

Here, we present an ODL-CIM model with squeezed reservoir modes. 
From the previous section, for a high success probability, 
the squeezed vacuum state with an anti-squeezed $\hat{X}$ component should be prepared for the reservoir related to the single mode loss, 
while a state with a squeezed $\hat{X}$ component should be prepared for the reservoir modes responsible for the mediating cavity loss. 
The Liouvillian of the squeezed reservoir modes is as follows \cite{Gardiner85}:
\begin{eqnarray}
\left.\frac{\partial \hat{\rho}}{\partial t}\right \vert_{C,sq} &=& 2n_s[\hat{a}_1,[\hat{\rho},\hat{a}_1^{\dagger}]]+2n_s[\hat{a}_2,[\hat{\rho},\hat{a}_2^{\dagger}]]\nonumber \\
&+& 2jn_j[\hat{a}_1-\hat{a}_2,[\hat{\rho},\hat{a}_1^{\dagger}-\hat{a}_2^{\dagger}]]\nonumber \\
&+& (m_s[\hat{a}_1,[\hat{a}_1,\hat{\rho}]]+m_s[\hat{a}_2,[\hat{a}_2,\hat{\rho}]]\nonumber \\
&-& jm_j[\hat{a}_1-\hat{a}_2,[\hat{a}_1-\hat{a}_2,\hat{\rho}]]+{\rm h.c.}).
\end{eqnarray}
Here, for the physical conditions of the reservoir mode to be satisfied, 
$n_s(1+n_s)\ge m_s^{2}$ and $n_j(1+n_j)\ge m_j^2$ must be satisfied. 
We will examine the reservoir mode with a minimum uncertainty product, 
where the above noise parameters are related to the phase sensitive gains $G_s$ and $G_j$ by 
\begin{eqnarray}
n_s &=& \frac{1}{4}\Bigl(G_s+\frac{1}{G_s}\Bigr)-\frac{1}{2}, \\
m_s &=& \frac{1}{4}\Bigl(G_s-\frac{1}{G_s}\Bigr), \\
n_j &=& \frac{1}{4}\Bigl(G_j+\frac{1}{G_j}\Bigr)-\frac{1}{2}, \\
m_j &=& \frac{1}{4}\Bigl(G_j-\frac{1}{G_j}\Bigr).
\end{eqnarray} 
In this case, the normalized correlation function is 
\begin{equation}
\mathcal{N}=1-\frac{\delta}{j}-\frac{2}{G_sG_j}\delta+O(\delta^2). 
\end{equation}
In the case of large $G_s G_j$, the normalized correlation function 
has an asymptotic value, $\mathcal{N}=1-\delta/j+O(\delta^2)$. 
A numerical simulation was performed with the Wigner SDEs: 
\begin{eqnarray}
\frac{d\alpha_1}{dt}&=& -(1+j)\alpha_1+p \alpha_1^*+j\alpha_2-g^2 |\alpha_1|^2\alpha_1 \nonumber \\
&+& \sqrt{\frac{1}{2}+n_s-m_s+g^2 |\alpha_1|^2}\xi_{C1} +\sqrt{2m_s}\xi_{R1} \nonumber \\
&+& \sqrt{j\Bigl(\frac{1}{2}+n_j-m_j\Bigr)}\xi_{C3}+i\sqrt{2jm_j}\xi_{R3}, \\
\frac{d\alpha_2}{dt}&=& -(1+j)\alpha_2+p \alpha_2^*+j\alpha_1-g^2 |\alpha_2|^2\alpha_2 \nonumber \\
&+& \sqrt{\frac{1}{2}+n_s-m_s+g^2 |\alpha_2|^2}\xi_{C2} +\sqrt{2m_s}\xi_{R2} \nonumber \\
&-& \sqrt{j\Bigl(\frac{1}{2}+n_j-m_j\Bigr)}\xi_{C3}-i\sqrt{2jm_j}\xi_{R3},
\end{eqnarray}
where $\langle \xi_{Ca}^*(t)\xi_{Cb}(t')\rangle=2\delta_{ab}\delta(t-t')$ and 
$\langle \xi_{Ra}(t)\xi_{Rb}(t')\rangle=\delta_{ab}\delta(t-t')$. 
The results for $G_s=G_j=10$ are shown in Fig. \ref{disc3}. 
The success probability is much larger than that of the MFB-CIM model and 
is higher than that of the ODL-CIM model with $n_{th}^s=10$ and $n_{th}^j=0$. 

\section{One-dimensional lattice}

\subsection{Steady-state fluctuation of ODL-CIM}

\begin{figure*}
\begin{center}
\includegraphics[width=14.0cm]{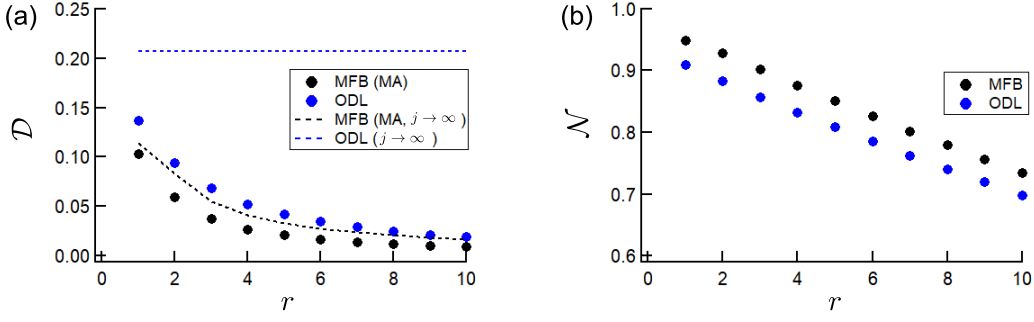}
\caption{Noise correlation in the 1D lattice. 
(a) Quantum discord $\mathcal{D}$ and (b) normalized correlation function $\mathcal{N}$ as a function of a distance $r$ with $p=0.999$ and $j=7/3$.}
\label{qd2}
\end{center}
\end{figure*}

In this section, we consider a ferromagnetic periodic one-dimensional lattice consisting of $N$-DOPOs 
represented by $\hat{a}_r (r=1,\cdots,N)$. 
First, we summarize the characteristics of the ODL-CIM model below the threshold \cite{Inui}. 
In the standing-wave model of an ODL-CIM \cite{Takata15,Maruo16}, 
the interaction is through the Liouvillian, 
\begin{equation}
\left.\frac{\partial \hat{\rho}}{\partial t}\right\vert_C=\frac{j}{2}\sum_{r=1}^N ([\hat{a}_r-\hat{a}_{r+1},\hat{\rho}(\hat{a}_r^{\dagger}-\hat{a}_{r+1}^{\dagger})]+{\rm h.c.}).
\end{equation}
Here, we assume periodicity: $\hat{a}_{N+1}=\hat{a}_1$. 
Assuming $p\sim 1$, the steady-state fluctuations of the canonical coordinates follow 
\begin{eqnarray}
\langle \Delta \hat{X}_1^2\rangle &=& \frac{1}{2}+\frac{p}{2\sqrt{2j(1-p)}}, \\
\langle \Delta \hat{X}_1 \Delta \hat{X}_{1+r} \rangle &=& \frac{p}{2\sqrt{2j(1-p)}}e^{-\sqrt{\frac{2(1-p)}{j}}r}. 
\end{eqnarray}
Assuming large $j$, the steady-state canonical momenta satisfy  
\begin{eqnarray}
\langle \Delta \hat{P}_1^2\rangle &=& \frac{1}{2}-\frac{p}{2\sqrt{2j(1+p)}}, \\
\langle \Delta \hat{P}_1 \Delta \hat{P}_{1+r} \rangle &=& -\frac{p}{2\sqrt{2j(1+p)}}e^{-\sqrt{\frac{2(1+p)}{j}}r}. 
\end{eqnarray}
The steady-state quantum discord of two DOPOs separated by a distance $r$ is calculated 
from the steady-state correlations. The results for $p=0.999$ and $j=7/3$ are shown in Fig. \ref{qd2} (a). 
The decay of the quantum discord is slow, because of the long-range correlation of the $\hat{X}$ components, 
whereas the $\hat{P}$ components have only short-range correlations. 
In the special limit $j\rightarrow \infty $ and $p\rightarrow 1$, 
the quantum discord of the ODL-CIM model is independent of $r$: $\mathcal{D}=f(\sqrt{2})+\frac{1}{2}\log \frac{1}{2}$. 
The normalized correlation function $\mathcal{N}$ for a pair of DOPOs separated by a distance $r$ is thus 
\begin{equation}
\mathcal{N}=1-\Bigl(\frac{r}{j}+1\Bigr)\sqrt{2j\delta}+O(\delta). 
\end{equation}
This function is plotted in Fig. \ref{qd2} (b) for $p=0.999$ and $j=7/3$. 

\subsection{Steady-state fluctuation of MFB-CIM (MA)}

Now let us consider the steady-state fluctuation of the MFB-CIM (MA) model. 
The Wigner SDE of the MFB-CIM (MA) is obtained as a special case of Eq. (\ref{sdema}) 
with $\tilde{J}_{r,r'}=\frac{1}{2}(\delta_{r,r'-1}+\delta_{r,r'+1})$. 
The coupling part is given as follows: 
\begin{eqnarray}
\left.\frac{d\alpha_r}{d t}\right\vert_C &=& -j\alpha_r+\frac{j}{2}({\rm Re}\alpha_{r-1}+{\rm Re}\alpha_{r+1})-\sqrt{\frac{j}{4}}\xi_{C1,r}\nonumber \\
&+& \frac{\sqrt{j}}{4}({\rm Re}\xi_{C1,r-1}+{\rm Re}\xi_{C1,r+1})+\sqrt{\frac{j}{4}}\xi_{C2,r}, 
\end{eqnarray}
where $\langle \xi_{Ca,r}^*(t)\xi_{Cb,r'}(t')\rangle=2\delta_{ab}\delta_{rr'}\delta(t-t')$. 
The over-all equation for $X$ containing a single mode part below the threshold is 
\begin{eqnarray}
\frac{dX_r}{dt} &=& -(1-p+j)X_r+\frac{j}{2}(X_{r-1}+X_{r+1})-\sqrt{\frac{j}{2}}\xi_{R1,r}\nonumber \\
&+& \sqrt{\frac{j}{8}}(\xi_{R1,r-1}+\xi_{R1,r+1})+\sqrt{1+\frac{j}{2}}\xi_{R2,r}, 
\end{eqnarray}
where $\langle \xi_{Ra,r}(t)\xi_{Rb,r'}(t')\rangle=\delta_{ab}\delta_{rr'}\delta(t-t')$. 
We take the Fourier transform of the canonical coordinates 
\begin{equation}
\tilde{X}_k=\frac{1}{\sqrt{N}}\sum_r X_r e^{i\theta_k r},
\end{equation}
where $\theta_k=\frac{2\pi}{N}k$. 
The Fourier components of the canonical coordinates satisfy the following SDE. 
\begin{eqnarray}
\frac{d\tilde{X}_k}{dt}&=& -(1-p+j(1-\cos\theta_k))\tilde{X}_k\nonumber \\
&-&\sqrt{\frac{j}{2}}(1-\cos\theta_k)\tilde{\xi}_{R1,k}+\sqrt{1+\frac{j}{2}}\tilde{\xi}_{R2,k}, 
\end{eqnarray}
where $\langle \tilde{\xi}_{Ra,k}\tilde{\xi}_{Rb,-k'}\rangle=\delta_{ab}\delta_{kk'}\delta(t-t')$. 
The steady-state fluctuations of the Fourier components are 
\begin{equation}
\langle \tilde{X}_k\tilde{X}_{-k}\rangle=\frac{1}{2}+\frac{p+\frac{j}{2}\cos^2 \theta_k}{2(1-p+j(1-\cos\theta_k))},
\end{equation}
which can be rewritten as 
\begin{equation}
\label{Xk2}
\langle \tilde{X}_k\tilde{X}_{-k}\rangle=\frac{1}{4}-\frac{1-p}{4j}-\frac{\cos\theta_k}{4}+\frac{p+\frac{(1-p+j)^2}{2j}}{2(1-p+j(1-\cos\theta_k))}. 
\end{equation}
The steady-state correlation function of the canonical coordinates is 
\begin{equation}
\langle \Delta \hat{X}_1 \Delta \hat{X}_{1+r}\rangle =\frac{1}{N}\sum_k \langle \tilde{X}_k\tilde{X}_{-k}\rangle e^{i\theta_k r}. 
\end{equation}
In Eq. (\ref{Xk2}), the first and the second terms of the R.H.S. contribute to the 
single mode fluctuation. The third term contributes to the correlation of a nearest-neighbor pair 
$\langle \Delta \hat{X}_1\Delta \hat{X}_2\rangle$, and the last term contributes to the correlation of a pair 
separated by a distance even more than 1. The correlation function is derived using the approximation $1-\cos \theta_k\sim \frac{\theta_k^2}{2}$, 
in the continuous limit $N\rightarrow \infty$: 
\begin{equation}
\frac{1}{N}\sum_k f(\theta_k) \rightarrow \frac{1}{2\pi}\int^{\pi}_{-\pi}f(\theta)d\theta,
\end{equation}
where the complex integration assumes a small $|1-p|/j$. The single mode fluctuation is 
\begin{equation}
\langle \Delta \hat{X}_1^2 \rangle=\frac{1}{4}-\frac{1-p}{4j}+\Bigl(p+\frac{(1-p+j)^2}{2j}\Bigr)\frac{1}{2\sqrt{2j(1-p)}}. 
\end{equation}
The correlation for a nearest-neighbor pair follows 
\begin{equation}
\langle \Delta \hat{X}_1 \Delta \hat{X}_{2} \rangle=-\frac{1}{8}+\Bigl(p+\frac{(1-p+j)^2}{2j}\Bigr)\frac{e^{-\sqrt{\frac{2(1-p)}{j}}}}{2\sqrt{2j(1-p)}}. 
\end{equation}
Here, the correlation function decreases by a factor $-1/8$. 
The correlation function for a pair separated by a distance $r>1$ is 
\begin{equation}
\langle \Delta \hat{X}_1 \Delta \hat{X}_{1+r} \rangle=\Bigl(p+\frac{(1-p+j)^2}{2j}\Bigr)\frac{e^{-\sqrt{\frac{2(1-p)}{j}}r}}{2\sqrt{2j(1-p)}}. 
\end{equation}
The quantum discord for the MFB-CIM (MA) model is calculated for $p=0.999$ and $j=7/3$. 
As shown in Fig. \ref{qd2} (a), this value is smaller than the quantum discord of the ODL-CIM model.  
In the limit $j\rightarrow \infty $ and $p\rightarrow 1$, 
the quantum discord for the MFB-CIM (MA) model depends on $r (>1)$: 
$\mathcal{D}=f(\sqrt{1+r})-f\Bigl(\sqrt{\frac{1+r}{2}}\Bigr)+\frac{1}{2}\log \frac{1}{2}$. 
When $r=1$, the quantum discord with $j\rightarrow \infty$ is the same as that for a two-DOPO MFB-CIM (MA) model: 
$\mathcal{D}=f(\sqrt{\frac{5}{2}})-f(\sqrt{\frac{5}{4}})+\frac{1}{2}\log \frac{1}{2}$.
The normalized correlation function $\mathcal{N}$ of the canonical coordinates is 
\begin{equation}
\mathcal{N}=1-\Bigl(\frac{r}{j}+\frac{1}{2+j}\Bigr)\sqrt{2j\delta}+O(\delta) 
\end{equation}
for $r>1$ and 
\begin{equation}
\mathcal{N}=1-\Bigl(\frac{1}{j}+\frac{3}{2(2+j)}\Bigr)\sqrt{2j\delta}+O(\delta) 
\end{equation}
for $r=1$. As shown in Fig. \ref{qd2} (b), the normalized correlation function of the MFB-CIM model 
is larger than that of the ODL-CIM model even when $r=1$, 
where the correlation function for the MFB-CIM (MA) model decreases by $-1/8$. 
The same results are obtained by the MFB-CIM (MI) model 
for ensemble-averaged values $\overline{\langle \overline{\Delta} \hat{X}_1^2\rangle}$ 
and $\overline{\langle \overline{\Delta} \hat{X}_1\rangle \langle \overline{\Delta}\hat{X}_{1+r}\rangle}$. 

\subsection{Success probability}

\begin{figure*}
\begin{center}
\includegraphics[width=14.0cm]{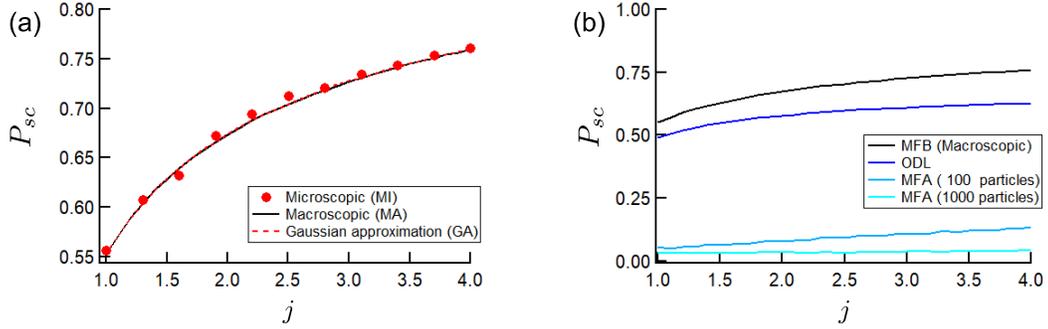}
\caption{Numerical success probability $P_{sc}$ of 1D lattice CIM ($N=6$) as a function of coupling coefficient $j$. 
(a) $P_{sc}$ of three MFB-CIM methods. (b) $P_{sc}$ of MFB-CIM, ODL-CIM, and MFA-CIM.}
\label{sp1d}
\end{center}
\end{figure*}

We calculated the success probability of a periodic one-dimensional lattice with $N=6$. 
A run was regarded as success when all the real parts of six Wigner amplitudes ${\rm Re}\alpha_r (r=1,\cdots,6)$ had the same sign in the final time step. 
Fig. \ref{sp1d} (a) presents the numerical success probability of the MFB-CIM models, i.e., (MI), (MA), and (GA). 
These three methods produce almost identical success probabilities. 
Fig. \ref{sp1d} (b) presents the numerical success probability of the MFB-CIM (MA), ODL-CIM, and MFA-CIM models with $10^2$ and $10^3$ Brownian particles. 
The MFB-CIM model has a higher success probability than that of the ODL-CIM model, 
whereas the ODL-CIM model with $j=7/3$ has a larger quantum discord than that of the MFB-CIM (MA) model. 
In the MFA-CIM model, noise correlation does not exist in the large Brownian particle number limit, 
and the success probability is much smaller. 
In the case of $10^3$ Brownian particles, the success probability is close to that of random-guess ($P_{sc}=\frac{1}{2^5}$). 
These results indicate that the success probabilities are more directly related to the normalized correlation function $\mathcal{N}$, 
than to quantum discord. 

\subsection{Several modified models}

\begin{figure*}
\begin{center}
\includegraphics[width=14.0cm]{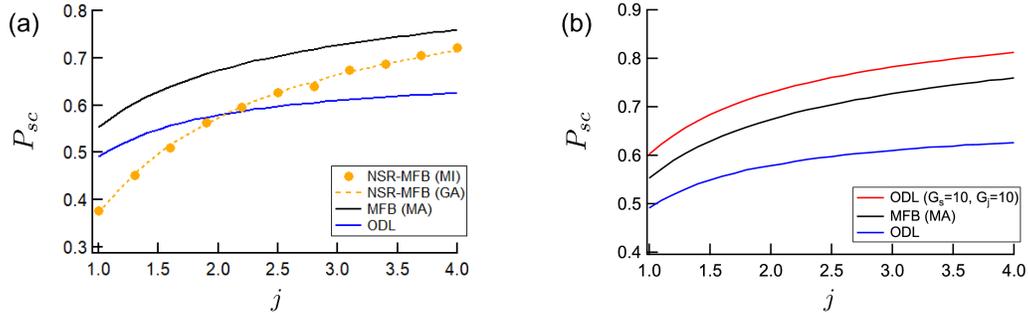}
\caption{Numerical success probability $P_{sc}$ of several 1D lattice CIM models ($N=6$) as a function of coupling coefficient $j$.
(a) Impact of measurement-induced state reduction on MFB-CIM. 
(b) Impact of squeezed reservoir on ODL-CIM. }
\label{disc4}
\end{center}
\end{figure*}

Here, we consider NSR-MFB-CIM models for a one-dimensional periodic lattice. 
The steady-state fluctuations of the canonical coordinates (before ensemble averaging) 
is the same as that of the mean field coupling $\langle \Delta \hat{X}_1^2\rangle=\frac{1}{2}+\frac{p}{2(1-p+j)}$. 
After ensemble averaging, the steady-state canonical coordinates have single site fluctuations, an $r=1$ correlation function, and an $r>1$ correlation function, as follows: 
\begin{eqnarray}
\overline{\langle \overline{\Delta} \hat{X}_1^2 \rangle} &=& \frac{1}{4}-\frac{1-p}{4j}+\frac{p}{2(1-p+j)}\nonumber \\
&+& \Bigl(\frac{(1-p+j)^2}{2j}\Bigr)\frac{1}{2\sqrt{2j(1-p)}}, 
\end{eqnarray}
\begin{equation}
\overline{\langle \overline{\Delta} \hat{X}_1 \rangle \langle \overline{\Delta} \hat{X}_{2} \rangle}=-\frac{1}{8}+\Bigl(\frac{(1-p+j)^2}{2j}\Bigr)\frac{e^{-\sqrt{\frac{2(1-p)}{j}}}}{2\sqrt{2j(1-p)}}, 
\end{equation}
\begin{equation}
\overline{\langle \overline{\Delta} \hat{X}_1 \rangle \langle \overline{\Delta} \hat{X}_{1+r} \rangle}=\Bigl(\frac{(1-p+j)^2}{2j}\Bigr)\frac{e^{-\sqrt{\frac{2(1-p)}{j}}r}}{2\sqrt{2j(1-p)}}. 
\end{equation}
Accordingly, the normalized correlation function for a pair with $r>1$ can be written as 
\begin{equation}
\mathcal{N}=1-\Bigl(\frac{r}{j}+\frac{j+2}{j^2}\Bigr)\sqrt{2j\delta}+O(\delta), 
\end{equation}
and for a nearest-neighbor pair ($r=1$) as 
\begin{equation}
\mathcal{N}=1-\Bigl(\frac{1}{j}+\frac{3j+4}{2j^2}\Bigr)\sqrt{2j\delta}+O(\delta).
\end{equation}
These values reach those of the ODL-CIM model at $j=2$ for $r>1$ and at $j \sim 2.35$ for $r=1$. 
Even for $r=1$, the coupling coefficient $j$ required to achieve the same $\mathcal{N}$ as the ODL-CIM model 
is smaller than that of the two-DOPO system ($j\sim 2.73$), 
because the ensemble-averaged single mode fluctuation is reduced in the one-dimensional NSR-MFB-CIM model. 
Fig. \ref{disc4} (a) shows the numerical success probabilities of the NSR-MFB-CIM (MI) and (GA) models. 
The success probabilities of the ODL-CIM and NSR-MFB-CIM models cross at $j\sim 2.05$, 
slightly beyond the point where the normalized correlation functions with $r>1$ coincide 
and before the point where those with $r=1$ coincide. 

Next, we consider an ODL-CIM model with squeezed reservoirs. 
The reservoir modes are squeezed by the following Liouvillian, 
\begin{eqnarray}
\left.\frac{\partial \hat{\rho}}{\partial t}\right\vert_{C,sq} &=& 2n_s\sum_r[\hat{a}_r,[\hat{\rho},\hat{a}_r^{\dagger}]]+m_s (\sum_r [\hat{a}_r,[\hat{a}_r,\hat{\rho}]]+{\rm h.c.}) \nonumber \\
&+& jn_j\sum_r [\hat{a}_r-\hat{a}_{r+1},[\hat{\rho},\hat{a}_r^{\dagger}-\hat{a}_{r+1}^{\dagger}]] \\
&-& \frac{j m_j}{2}(\sum_r[\hat{a}_r-\hat{a}_{r+1},[\hat{a}_r-\hat{a}_{r+1},\hat{\rho}]]+{\rm h.c.}). \nonumber
\end{eqnarray}
We assume the reservoir with minimum uncertainty state 
described by phase sensitive gain ($G_s=1+2(n_s+m_s),G_j=1+2(n_j+m_j)$). 
The normalized correlation function for a pair separated by a distance $r$ is 
\begin{equation}
\mathcal{N}=1-\Bigl(\frac{r}{j}+\frac{1}{G_s G_j}\Bigr)\sqrt{2j\delta}+O(\delta). 
\end{equation}
$\mathcal{N}=1-r\sqrt{2\delta/j}+O(\delta)$ in the limit of large $G_s G_j$. 
The numerical success probability with $G_s=G_j=10$ is shown in Fig. \ref{disc4} (b). 
The success probability of the ODL-CIM model with squeezed reservoirs exceeded that of the MFB-CIM model. 

\section{Summary}

We compared the noise correlations and 
success probabilities of the MFB-CIM, ODL-CIM and MFA-CIM models 
for the cases of two DOPOs and of a periodic one-dimensional lattice. 
For two DOPOs, a numerical simulation was performed in parameter spaces 
where the ODL-CIM model satisfies the entanglement criterion, 
and has a larger quantum discord than that of the MFB-CIM model. 
We note the surprising fact that the MFB-CIM model 
has a larger success probability than the ODL-CIM model. 
In fact, the MFB-CIM model has a larger normalized correlation function of the canonical coordinates $\hat{X}$ than 
the ODL-CIM model has in those parameter spaces. 
These results can be understood by the following argument: 
the canonical momentum $\hat{P}$, which is included in the entanglement and quantum discord, 
but not in the normalized correlation function,
does not contribute directly to the computation process in a CIM. 
The CIM works only through the canonical coordinate $\hat{X}$. 
We showed that the normalized correlation function predicts the point 
where the success probabilities of two different CIMs using different coupling schemes cross over. 
We analyzed the ODL-CIM with squeezed reservoir modes 
and showed that they have higher success probabilities than MFB-CIM does. 
We also showed that the normalized correlation function is a useful metric 
to predict the success probability in a one-dimensional lattice of CIMs. 

\

The numerical method in the paper was used in Ref.\cite{Kako20}.

\appendix

\section{Gaussian homodyne-measurement theory}

Here, we discuss the theory of measurement-induced state reduction. 
In Fig. \ref{model}, we assume that the canonical coordinate of the reflected mode $\hat{X}_{R,r}$ is measured 
and that a random measurement result $X_{M,r}$ is obtained. 
The difference of the measured value from the mean value is denoted as $d_r:=X_{M,r}-\langle \hat{X}_{R,r}\rangle$. 
From Gaussian homodyne measurement theory \cite{Eisert02}, 
the mean amplitude is shifted and the variance of the transmitted mode is reduced respectively in reaction to a measurement: 
\begin{eqnarray}
\langle \hat{X}_{T,r}'\rangle &=& \langle \hat{X}_{T,r}\rangle+\frac{\langle \Delta \hat{X}_{T,r}\Delta \hat{X}_{R,r}\rangle}{\frac{1}{2}+\langle :\Delta \hat{X}_{R,r}^2:\rangle}d_r, \\
\langle :\Delta \hat{X}_{T,r}'^2:\rangle &=& \langle:\Delta \hat{X}_{T,r}^2:\rangle-\frac{\langle \Delta \hat{X}_{T,r}\Delta \hat{X}_{R,r}\rangle^2}{\frac{1}{2}+\langle :\Delta \hat{X}_{R,r}^2:\rangle}.
\end{eqnarray}
In the above two equations, the variance of the reflected component in the 
denominator $\langle :\Delta \hat{X}_{R,r}^2:\rangle=R_B\langle :\Delta \hat{X}_r^2:\rangle$ is negligible 
when $R_B=j\Delta t$ is sufficiently smaller than one. 
Note that $\langle:\Delta \hat{X}^2:\rangle$ is a normally ordered variance, 
where the contribution of the vacuum fluctuation has been removed. 
When $R_B \ll 1$, from $\langle \Delta \hat{X}_{T,r}\Delta \hat{X}_{R.r}\rangle=\sqrt{R_B}\langle :\Delta \hat{X}_{r}^2:\rangle$, 
the variance reduction follows: 
\begin{equation}
\label{nr1}
\langle :\Delta \hat{X}_{T,r}'^2:\rangle=\langle:\Delta \hat{X}_{T,r}^2:\rangle-2j\Delta t \langle :\Delta \hat{X}_{r}^2:\rangle^2. 
\end{equation}
Next, we consider the shift of the mean amplitude. 
In the canonical coordinate of the reflected mode, 
we assume that the fluctuation of $X_r$ is sufficiently smaller than the vacuum noise. 
The reflected mode is $X_{R,r}\sim \sqrt{R_B}\langle X_r\rangle -\sqrt{2(1-R_B)}{\rm Re}f_{1r}$.
Therefore, $d_r$ for each $r$ is a real number selected from the distribution of 
$-\sqrt{2(1-R_B)}{\rm Re} f_{1r}$, where $\langle {\rm Re} f_{1r} {\rm Re}f_{1r'}\rangle=\frac{1}{4}\delta_{r,r'}$. 
Following the notation in Refs. \cite{Wiseman93,Shoji17} for measurement noise, 
$d_r$ is written as $d_r=\sqrt{\frac{\Delta t}{2}}w_{R,r}$, 
where $w_{R,r}$ follows $\overline{w_{R,r}(t)w_{R,r'}(t')}=\delta_{r,r'}\delta(t-t')$ under the ensemble averaging. 
The shift in the mean amplitude in the limit $R_B\ll 1$ is 
\begin{equation}
\label{nr2}
\langle \hat{X}_{T,r}'\rangle=\langle \hat{X}_{T,r}\rangle+\sqrt{2j}\langle :\Delta \hat{X}_{r}^2:\rangle w_{R,r} \Delta t.
\end{equation}
Relations (\ref{nr1}) and (\ref{nr2}) represent the measurement-induced state reduction 
and are incorporated in the Wigner SDE (Eq. (\ref{Xms})). 

\section{Effect of saturation parameter $g^2$}

\begin{figure}[h]
\begin{center}
\includegraphics[width=7.0cm]{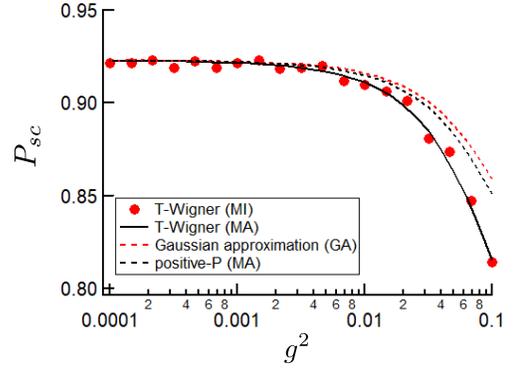}
\caption{Numerical success probability $P_{sc}$ of two-site MFB-CIM by truncated Wigner and positive-$P$ SDE 
as a function of saturation coefficient $g^2$ for $j=2$.}
\label{fa}
\end{center}
\end{figure}

The positive-$P$ simulation \cite{Takata15,Inui} does not require truncation based on the assumption $g^2 \ll 1$ 
and is expected to be exact for a larger $g^2$. 
Here, we compared the results of a simulation based on truncated Wigner theory with 
those of positive-$P$ SDE theory for a MFB-CIM model consisting of two DOPOs. 
Positive-$P$ theory expands the density matrix \cite{Drummond80} 
as $\hat{\rho}=\int P(\alpha,\alpha^{\dagger}) \frac{|\alpha \rangle \langle \alpha^{\dagger *}|}{\langle \alpha^{\dagger *}|\alpha\rangle }d^2\alpha d^2\alpha^{\dagger}$. 
The positive-$P$ SDEs for a macroscopic model are 
\begin{eqnarray}
\frac{d\alpha_r}{d t} &=& -(1+j)\alpha_r+p\alpha_r^{\dagger}-g^2\alpha_r^{\dagger}\alpha_r^2+\sqrt{p-g^2\alpha_r^2}\xi_{R1,r} \nonumber \\
&+& \sum_{r'}\tilde{J}_{r,r'}\Bigl(j\frac{\alpha_{r'}+\alpha_{r'}^{\dagger}}{2}+\sqrt{\frac{j}{4}}\xi_{R3,r'}\Bigr),\\
\frac{d\alpha_r^{\dagger}}{d t} &=& -(1+j)\alpha_r^{\dagger}+p\alpha_r-g^2\alpha_r^{\dagger 2}\alpha_r+\sqrt{p-g^2\alpha_r^{\dagger 2}}\xi_{R2,r} \nonumber \\
&+& \sum_{r'}\tilde{J}_{r,r'}\Bigl(j\frac{\alpha_{r'}+\alpha_{r'}^{\dagger}}{2}+\sqrt{\frac{j}{4}}\xi_{R3,r'}\Bigr), 
\end{eqnarray}
where $\langle \xi_{Ra,r}(t)\xi_{Rb,r'}(t')\rangle=\delta_{ab}\delta_{rr'}\delta(t-t')$. 
The amplitudes $(\alpha_r,\alpha_r^{\dagger})$ in the positive-$P$ expansion do not contain the vacuum fluctuation, 
but the Wigner amplitude $\alpha_r$ does. 
For a fair comparison,  
we judged the success using the signs of ${\rm Re}\frac{\alpha_r+\alpha_r^{\dagger}}{\sqrt{2}}+\frac{N_r}{\sqrt{2}}$ in the positive-$P$ simulation, 
where $N_r$ is a normal random variable. 
When the signs were same for two DOPOs, we judged that the run was success. 
In Fig. \ref{fa}, the success probability of the MFB-CIM models (MA, MI, and GA) with $j=2$ 
were calculated with the truncated Wigner (T-Wigner) simulation. 
A positive-$P$ simulation was also carried out on the MA model. 
The number of runs was $10^4$ for the MI model and $10^7$ for the other methods. 
The resulting success probabilities of the two-site system were smaller for large $g^2$, 
and the positive-$P$ simulation produced slightly larger success probabilities than the truncated Wigner simulation did. 
The normalized correlation function of $\hat{X}$ accounts for the decreasing success probability for larger $g^2$. 
When $g^2$ is large, the role of the vacuum fluctuation in $\langle \Delta\hat{X}^2\rangle$ 
appearing in the denominator of the normalized correlation function is larger, 
and the normalized correlation function at the threshold is smaller. 
For small $g^2$, such as $g^2=10^{-4}$ employed in the main text, 
Wigner and positive-$P$ simulation produce almost identical success probabilities. 
However, for large $g^2$, the truncated Wigner approach produces inaccurate results compared with positive-$P$ theory 
due to the truncation assuming small $g^2$ values. 
The results for the GA model differ from those of other truncated Wigner models (MA, and MI) 
because of the additional truncation used to separate the $\hat{P}$ components from the $\hat{X}$ components. 

\begin{thebibliography}{99}

\bibitem{Wang13} Z. Wang, A. Marandi, K. Wen, R. L. Byer, and Y. Yamamoto, Phys. Rev. A 88,063853 (2013). 
\bibitem{Marandi14} A. Marandi, Z. Wang, K. Takata, R. L. Byer, and Y. Yamamoto, Nature Photon. 8, 937 (2014).
\bibitem{Takata16} K. Takata, et al., Sci. Rep. 6, 34089 (2016). 
\bibitem{Inagaki16} T. Inagaki, K. Inaba, R. Hamerly, K. Inoue, Y. Yamamoto, and H. Takesue, Nature Photon. 10, 415 (2016).

\bibitem{McMahon16} P. L. McMahon, A. Marandi, Y. Haribara, R. Hamerly, C. Langrock, S. Tamate, and R. L. Byer, Science 354, 614 (2016).
\bibitem{Inagaki16b} T. Inagaki, et al., Science 354, 603 (2016).
\bibitem{Hamerly19} R. Hamerly, et al., Sci. Adv. 5, eaau0823 (2019). 

\bibitem{Wolinsky88} M. Wolinsky, and H. J. Carmichael, Phys. Rev. Lett. 60, 1836 (1988).
\bibitem{Kinsler91} P. Kinsler, and P. D. Drummond, Phys. Rev. A 43, 6194 (1991). 


\bibitem{Duan00} L.M.Duan, G.Giedke, J.I.Cirac, and P.Zoller, Phys.Rev.Lett.84,2722(2000).
\bibitem{Takata15} K. Takata, A. Marandi, and Y. Yamamoto, Phys. Rev. A 92,043821 (2015).
\bibitem{Maruo16} D. Maruo, S. Utsunomiya, and Y. Yamamoto, Phys. Scr. 91,083010 (2016).

\bibitem{Inui} Y.Inui, and Y.Yamamoto, arXiv:1905.12348v2 (2019). 

\bibitem{Cahill69} K.E.Cahill, and R.J.Glauber, Phys.Rev.177,1882(1969). 
\bibitem{Walls} D.F.Walls, and G.J.Milburn, "Quantum Optics", Springer (2007).
\bibitem{Corney03} J.F.Corney, and P.D.Drummond, Phys.Rev.A 68,063822(2003).

\bibitem{Braginsky} B.V.Braginsky, V.B.Braginskii, and F.Y.Khalili, "Quantum measurement", Cambridge(1995). 

\bibitem{Wiseman93} H.M.Wiseman and G.J.Milburn, Phys.Rev.Lett. 70,548 (1993).

\bibitem{Haribara15} Y.Haribara, Y.Yamamoto, K,Karawabayashi, and S.Utsunomiya, arXiv:1501.07030v1(2015).
\bibitem{Haribara17} Y.Haribara, H.Ishikawa, S.Utsunomiya, K.Aihara, and Y.Yamamoto, Quantum Science and Technol. 2, 044002(2017).

\bibitem{Shoji17} T. Shoji, K. Aihara, and Y. Yamamoto, Phys. Rev. A 96, 053833 (2017).

\bibitem{Eisert02} J. Eisert, S. Scheel, and M. B. Plenio, Phys. Rev. Lett. 89, 137903 (2002). 

\bibitem{Wiseman93b} H.M.Wiseman and G.J.Milburn, Phys.Rev.A 47,642 (1993).

\bibitem{Giorda10} P. Giorda and M. G. A. Paris, Phys. Rev. Lett. 105, 020503 (2010).
\bibitem{Adesso10} G. Adesso and A. Datta, Phys. Rev. Lett. 105, 030501 (2010).



\bibitem{Gardiner85} C.W.Gardiner, and M.J.Collett, Phys.Rev.A 31, 3761 (1985).

\bibitem{Drummond80} P. D. Drummond and C. W. Gardiner, J. Phys. A 13, 2353 (1980).

\bibitem{Kako20} S. Kako, T. Leleu, Y. Inui, F. Khoyratee, S. Reifenstein, and Y. Yamamoto, Adv. Quantum Technol. 2000045 (2020). 



\end{thebibliography}
\end{document}